\documentclass[
nobibnotes,
amsmath,amssymb,aps,
prl,
twocolumn,
]{revtex4-2}
\usepackage{xcolor}
\usepackage{enumitem}
\usepackage{float}
\usepackage{color}
\usepackage{graphicx}
\usepackage{dcolumn}
\usepackage{bm}
\usepackage{natbib}
\usepackage{amsfonts,bbm}

\def \dd {\bm{d}}
\def \xx {\bm{x}}
\def \XX {\bm{X}}

\def \Xmu {X_{\mu}}
\def \kB {k_{\rm B}}

\newcommand{\dStot}{\Delta S_{tot}}

\newcommand{\bra}[1]{\left\langle #1 \right\rangle}

\begin{document}

\title{Quantitative analysis of non-equilibrium systems from short-time experimental data}
\author{Sreekanth K Manikandan\textsuperscript{1,}\textsuperscript{4}}
\author{Subhrokoli Ghosh\textsuperscript{2,}\textsuperscript{4}}
\author{Avijit Kundu\textsuperscript{2}}
\author{Biswajit Das\textsuperscript{2}}
\author{Vipin Agrawal\textsuperscript{1,}\textsuperscript{3}}
\author{Dhrubaditya Mitra\textsuperscript{1,}\textsuperscript{3}}
\email{dhruba.mitra@gmail.com}
\author{Ayan Banerjee\textsuperscript{2}}
\email{ayan@iiserkol.ac.in}
 \author{Supriya Krishnamurthy\textsuperscript{3}}%
 \email{supriya@fysik.su.se}
\affiliation{\textsuperscript{1}NORDITA, KTH Royal institute of technology and Stockholm university, Stockholm.}
\affiliation{\textsuperscript{2}Department of Physical Sciences, IISER Kolkata, India}
\affiliation{\textsuperscript{3}Department of Physics, Stockholm University, SE-10691 Stockholm, Sweden}%
\affiliation{\textsuperscript{4}These authors contributed equally: Sreekanth K Manikandan and Subhrokoli Ghosh }%

\date{\today}

\begin{abstract}
 We provide a minimal strategy for the quantitative analysis of a large class of non-equilibrium  systems in a {statistically} steady state using the short-time Thermodynamic Uncertainty Relation (TUR).
From short-time trajectory data obtained from experiments, we demonstrate how we can simultaneously infer quantitatively, both the 
thermodynamic force field acting on the system, as well as the (potentially exact) rate of entropy production. We benchmark this scheme first for an experimental study of a colloidal particle system where exact analytical results are known, before applying it to the case of a colloidal particle in a hydrodynamical flow field, where neither analytical nor numerical results are available.   In this latter case, we build an effective model of the system based on our results. In both cases, we also demonstrate that  our results match with those obtained from another recently introduced scheme \cite{frishman:sfi}.

\end{abstract}

\maketitle
$ $\\
Non-equilibrium thermodynamics at microscopic length scales is dominated by a fascinating range of  phenomena \cite{bustamante_phystoday_2005}, where thermal fluctuations play a crucial role. 
These phenomena can now be observed in great detail experimentally, due to the availability and scope of current microscopic manipulation techniques.
The interpretation and quantitative analysis of the experimentally available data is however lagging behind these advances, mostly due to the fact that the vast majority of these systems are  too complicated to model without making several approximations, despite having far fewer degrees of freedom than their macroscopic counterparts. Even when it is possible to build such simplified models, these are  still usually too complicated to solve except sometimes by numerical analysis of specific systems, which however lack general insights. 
There could also be other factors making the system hard to solve, such as the  presence of  a background flow, for which the spatial dependence of the flow velocity needs to be known by means of solving the corresponding Navier-Stokes equation; usually a difficult task,  
especially for unsteady flows.  In the face of all these challenges, a relevant question is whether it is at all possible to
gain any precise quantitative information about a complex non-equilibrium system directly from experimental data, bypassing the first step of either having a known model to compare with or building in simplifying assumptions about the system. 

Not surprisingly, this question has aroused a lot of recent interest.
Broadly speaking, measurements from experiments can be used to
obtain general information about the system, such
as identifying that detailed balance is broken and hence the system is out-of-equilibrium \cite{battle2016broken,seara2018entropy,expmetric} (not always obvious for microscopic systems such as at the cellular level), or to obtain more
specific properties of the system such as the rate of dissipation of energy (equivalently the rate of entropy production) \cite{harada2005equality,noninvasive, muy2013non, parrondo2002energetics,roldan2010estimating,Gingrich:qua,gnesotto:lnb,tusch2014energy}, the average phase-space velocity field \cite{zia2007probability, battle2016broken, frishman:sfi} related to the so-called thermodynamic force field \cite{qian2001mesoscopic,van2010three} or the microscopic forces driving the system \cite{Volpe:Forma, frishman:sfi, gnesotto:lnb}. The motivation for such studies is that if quantitative information about the system can be directly obtained from experimentally observed quantities, then this understanding can be used for building more realistic and experimentally validated models of the system of interest \cite{battle2016broken,turlier2016equilibrium,wan2018time}.

A very informative quantity about a non-equilibrium system is the 
rate of entropy production. 
This quantity not only
signals - when it is non-zero - that the system is out of equilibrium, but also provides a quantitative measure of how far from equilibrium a system is and the irreversibility of the dynamics \cite{seifert2008stochastic,jarzynski2011equalities,Seifert:2012stf}. 
In the context of microscopic machines \cite{Manikandan:EF}, a 
quantification of the amount of energy dissipated directly provides information 
about engine efficiencies \cite{ martinez2017colloidal,Verley:2014uce,Verley:2014ute} and prescriptions for
obtaining optimal operating conditions \cite{paneru2018optimal}. The value of the entropy production rate can also be used to obtain
information-theoretic quantities of interest \cite{parrondo2015thermodynamics}, or even
information about hidden degrees of freedom \cite{martinez2019inferring}. 
 The entropy production rate is also a very robust quantity to measure from the experimental point of view, since it is not so strongly affected by conversion-factor errors in measuring particle positions, as we remark later.

The entropy production rate can be obtained directly from experimental data, at least for systems where it is understood that the underlying dynamics is Markovian, by several means.  These include utilizing the Harada-Sasa equality \cite{harada2005equality}
which involves a spectral analysis of trajectory data \cite{oocites,measuring}, determining the average steady state current and steady-state probability distribution from the data \cite{noninvasive}, determining the time-irreversibility of the dynamics \cite{sekimoto1997kinetic,sekimoto1998langevin,Seifert:2005epa,Seifert:2012stf,maes2003time,gaspard2004time} and relatedly determining estimators for the ratio of forward and backward processes directly from the data \cite{roldan2010estimating,andrieux2008thermodynamic,EpTA}.
Recent approaches \cite{frishman:sfi, gnesotto:lnb}
also advocate inferring first the microscopic force field from which the entropy production rate can be {deduced}. 

An alternative strategy to direct estimation, is to set lower bounds on the entropy production rate \cite{kawai2007dissipation, Blythe2008, Vaikuntanathan2009, Muy2013,PhysRevLett.114.158101} by measuring experimentally accessible quantities. 
One class of these bounds, for example those based on the thermodynamic uncertainty relation (TUR) \cite{PhysRevLett.114.158101,gingrich:pft,Ginrich:dba,Seifert:inf,horowitz:tuc}, have been further developed into variational {\em inference} schemes which translate the task of identifying entropy production to an optimization problem over the space of a single projected fluctuating current in the system \cite{Gingrich:qua,manikandan2019inferring,Shun:eem,van:epe}. Recently, a similar variational scheme using neural networks was also proposed \cite{kim2020learning}. { As compared to some of the other trajectory-based entropy estimation methods, these inference schemes do not involve the estimation of probability distributions over the phase-space}. Rather they usually only involve means and variances of measured currents. Hence they are known to work better in higher dimensional systems \cite{Gingrich:qua}.
In addition, it is proven that such an optimization problem gives the exact value of the entropy production rate in a steady state as well as the exact value of the thermodynamic force field in the phase space of the degrees of freedom we can measure, if short-time currents are used \cite{manikandan2019inferring,Shun:eem,van:epe,kim2020learning}. However, these methods have not yet been tested against experimental data to the best of our knowledge.

 Here we test the Short-time TUR scheme against the challenges posed by experimental setups involving
colloidal particles in time-varying potentials with (possible) background flows. 
In order to benchmark the scheme, we first test the scheme 
in a setup where the entropy production rate of the system can be analytically predicted for any set of parameters.
For this set up, we test our predictions against both analytical results as well as another  recently  proposed numerical scheme, namely stochastic force inference (SFI) \cite{frishman:sfi}. 
After this benchmarking exercise, we apply our scheme to a modified system for which the underlying model is both unknown and hard to estimate. Though there is no theoretical value to compare with in this case, the short-time TUR's predictions even here, are again in perfect agreement with that predicted by the SFI technique \cite{frishman:sfi}. These results provide a motivation for modelling this system in terms of
coupled Langevin equations with two free parameters. We demonstrate that such a model does indeed capture 
the experimental observations, hence demonstrating the usefulness of these schemes in modeling complex scenarios.

\section{Model}
The results we demonstrate here apply to systems with continuous state-space but a finite-number of degrees of freedom, described
by overdamped Langevin equations of the type
 \begin{align}
 \dot{X}_{\mu}(t)=F_{\mu}[\XX(t)]+G_{\mu \nu}[\XX(t)]\cdot\xi_{\nu}\/,
 \label{eq:Langevin}
 \end{align}
 Here $\mu = 1,\ldots, \mathbbm{d}$ is the number of degrees of freedom of the system and we use $\cdot$ to refer to the Ito convention.
 $F_{\mu}(\XX)$ is a function of $\XX$, but not an explicit function of time $t$,
 $\xi_{\mu}$ is $\mathbbm{d}-$ dimensional white-in-time noise such that
 $\bra{\xi_{\mu}(t)\xi_{\nu}(t^{\prime})} = \delta_{\mu\nu}\delta(t-t^{\prime})$,
 where $\bra{\cdot}$ denotes averaging over the statistics of the noise. 
 The corresponding Fokker--Planck equation for the probability distribution function
 $P$ is given by:
 \begin{subequations}
   \begin{align}
   \partial_t P &= -\partial_{\mu} J_{\mu} \quad\/,\\
   J_{\mu} &\equiv F_{\mu} P - D_{\mu \nu}\partial_{\nu} P\/,& D_{\mu \nu}&=\frac{1}{2}G_{\mu \alpha}G_{\alpha \nu},
\end{align}
 \end{subequations}
 \textcolor{black}{where the repeated indices are summed over.}
 In the steady state $\partial_t P = 0$.
 The total rate of entropy production  $\sigma$ can be obtained as~\cite{Seifert:2005epa,noninvasive}, 
\begin{subequations}
 \begin{align}
   \sigma &=\int d\XX\;  \mathcal{F}_{\mu}J_{\mu}  \/\quad{\rm where}\label{eq:sigma}\\
   \mathcal{F}_{\mu} &\equiv \frac{D^{-1}_{\mu \nu}J_{\nu}}{P} \label{eq:mF}
 \end{align}
\end{subequations}
is called the thermodynamic force field \cite{Gingrich:qua}.
Overdamped Langevin equations are excellent descriptions for colloidal particle systems. 
Even for systems where the Langevin equation is not known,
the fact that such a description exists in principle is all that is needed in order to apply Eq.~\ref{eq:sigma} and obtain
$\sigma $  by determining the current and steady-state probability density directly from the
time-series data \cite{Gingrich:qua,noninvasive}.
Another approach is to first infer the terms in the Langevin equation,  $F_{\mu}$ and $D$ \cite{frishman:sfi, gnesotto:lnb}
and use Eq.~\ref{eq:sigma} to obtain $\sigma $.
These methods can be applied directly on data obtained from tracking the system or  even by using
tracking-free methods in image space \cite{gnesotto:lnb}.

\section{Results}

In this paper, we demonstrate an alternative method for the simultaneous determination of
both the entropy production rate as well as the thermodynamic force field $\mathcal{F}_{\mu} $
from experimental data, using the recently introduced short-time thermodynamic inference
relation \cite{manikandan2019inferring,Shun:eem,van:epe}.
Our method is built on an exact result obtained in \cite{manikandan2019inferring,Shun:eem,van:epe}:
\begin{align}
\label{eq:unc}
   \sigma =\max_J\left[ \frac{2 \kB \bra{J} ^2 }{\Delta t \text{Var}(J)} \right]\/,
\end{align}
where $\kB$ is the Boltzmann constant and $J$ is  a weighted scalar current constructed from the non-equilibrium stationary state  as shown below. The notation $\langle \cdot \rangle$ stands for an ensemble average. \textcolor{black}{The current that maximizes the term within the square brackets is $J\propto\Delta S_{tot}$. Here $\Delta t$ is the short time interval over which the mean and variance of the current is evaluated \cite{manikandan2019inferring}. In this work, it also coincides with the sampling rate of the trajectory}. 
As for the ordinary TUR ~\cite{fischer2020free}, our result too holds 
for any $\XX$ that is even under time reversal. The  equality in \eqref{eq:unc} holds only when $\XX$ includes {\bf all} degrees of freedom of the system. If not, then the RHS of \eqref{eq:unc} gives a lower bound. The proof presented for Eq.\ \eqref{eq:unc} in \cite{manikandan2019inferring} was based on  exact results for non-trivial models. \textcolor{black}{It was shown that Eq.\ \eqref{eq:unc} is a consequence of \textcolor{black}{fluctuations of $\Delta S_{tot}$ becoming Gaussian}, in the $\Delta t \rightarrow 0$ limit. Later in \cite{Shun:eem} and \cite{van:epe}, Eq.\ \eqref{eq:unc} was rigorously proved for overdamped diffusive processes. }

Let us now discretize $\XX$ in time with time interval $\Delta t$:
$\Xmu^{0}\cdots \Xmu^{\rm{j}}\cdots \Xmu^{\rm{N}}$. 
We use latin indices as superscripts for the discrete time labels
{and Einstein summation convention is applied to the greek indices}. 
For a given function $\dd(\XX)$ we can now define a time-discretised scalar  function constructed from the steady state current, 
\begin{equation}
  J^{\rm k} = d_{\mu}\left( \frac{\XX^{\rm k}+\XX^{\rm{k+1}}}{2} \right) 
  \left(\Xmu^{\rm{k+1}}- \Xmu^{{\rm k}} \right)
  \label{eq:JJ}
\end{equation}
{\color{black} Any such current, when substituted in the expression inside the square brackets of Eq. \ref{eq:unc} can be shown to give a lower bound $\sigma_L$ which is $ \leq \sigma$.  In addition, for a special value of $d = d^{\ast}$, $J\propto\dStot$ and $\sigma_L=\sigma $.}
The algorithm we use, which obtains this $\mathbbm{d}^{\ast}$ and $\sigma$ through a maximisation procedure is as follows:
\begin{enumerate}
\item We first obtain a time-series of experimental data: $\XX^{\rm k}$.
\item To be able to perform the maximisation we use a set of basis functions $\psi_m(\XX)$, $m=1,\ldots, M$, in the space spanned by $\XX$ such that
  \begin{equation}
    {\bm d}(\XX) = \sum_{m=1}^{M}{\bm w}_m\psi_m(\XX)\/,
    \label{eq:psi}
  \end{equation}
{\color{black} where ${\bm w}_m \in \mathbb{R}^\mathbbm{d} $ and are the parameters to be optimised.} 
 We use two sets of basis functions: Gaussian and linear and generate all our results in both these bases, for comparison. 
  \item Maximize Eq.\ \eqref{eq:unc} to obtain $\sigma$.  This maximization is done using a numerical optimizer: We start with an initial guess for $w_m$, calculate the time-series $J^k$,
  construct the function within the
  square brackets in \eqref{eq:unc}
  and then maximise over ${\bm w}_m$ to obtain $\sigma$
  and also the set of values $w^{\ast}_m$ such that 
  ${\bm d}^{\ast} = \sum_{m=1}^{M}{\bm w}^{\ast}_m\psi_m(\XX)$   maximises Eq.~\eqref{eq:unc}.
The maximising current $J^{\ast}$ is constructed from $d^{\ast}$ using Eq.\ \eqref{eq:JJ} and in addition can be shown to be proportional to $\dStot $ \cite{manikandan2019inferring}. 
\end{enumerate}
Furthermore, the thermodynamic force is proportional to  ${\bm d}^{\ast}$ that maximises \eqref{eq:unc} \cite{manikandan2019inferring,Shun:eem,van:epe}, i.e., 
\begin{equation}
\label{eq:optf}
  \mathcal{F} \propto {\bm d}^{\ast}
\end{equation}
 Hence, by solving an optimization problem, where the RHS of Eq.\ \eqref{eq:unc} is
maximized in the space of all currents we can obtain $\sigma$ as the optimal value as well as
its conjugate thermodynamic force field, $\mathcal{F}=c\;\textit{\textbf{d}}^{\ast}$  where the proportionality constant  can be fixed by using  Var$(J^{\ast}) = 2 \langle {J^\ast} \rangle $ at $\Delta t \rightarrow 0$ \cite{manikandan2019inferring} as, $c = \frac{2 \langle J^\ast \rangle}{\text{Var}(J^\ast)}$. 

 \textcolor{black}{We note that, for any set of basis functions $\psi_m(\XX)$, $m=1,\ldots, M$ which give an adequate 
 representation of ${\bm d}(\XX)$, an analytic solution to the maximization problem is known \cite{van:epe}. This solution gives a deterministic estimate of $\sigma$ as,
\begin{align}
    \sigma = \frac{2 \bar{\psi}_{k} \;(\Xi ^{-1})_{k,l}\; \bar{\psi}_{l}}{\Delta t},
\end{align}
where $\bar{\psi}_{k} = \langle\psi_k\rangle $ and $\Xi_{k,l }= \langle\psi_k\psi_l\rangle-\bar{\psi}_{k}\bar{\psi}_{l}$. Further, the optimal coefficients can be directly computed without any optimisation as
\begin{align}
    \omega^\ast_k =  \frac{(\Xi ^{-1})_{k,l}\; \bar{\psi}_{l}}{(\bar{\psi}_{k} \;(\Xi ^{-1})_{k,l}\; \bar{\psi}_{l})}.
\end{align}
Repeated indices are summed over as before.   Numerically, this involves inversion of the matrix $\Xi$. 
On the one hand, if $\mathbbm{d},\;N$ and $M$ are not very large, this deterministic scheme is faster compared to a numerical   optimization algorithm, and does not get stuck in local maxima. On the other hand, numerical optimization schemes can in principle simultaneously handle the  optimisation of parameters of the basis functions. This is discussed in some detail in \cite{Shun:eem}. In addition, numerical optimization schemes have also been extended to systems driven in a time dependent manner \cite{otsubo2020estimating}, where it is as yet unclear how the deterministic scheme will perform.} 
 
 In this work, we implement the numerical optimization scheme using a \textit{particle-swarm} optimizer. We provide a brief introduction to the algorithm in the Methods section and also study its convergence properties in Figs. $9-13$. We note that Refs \cite{Shun:eem,van:epe} have already demonstrated the feasibility of the scheme described here with numerical data. Here we test this scheme instead on  controlled experimental setups. 

\subsubsection{Colloidal particle in a stochastically shaken trap}
To test the inference scheme we first apply it to an experimental problem for
which the rate of entropy production is known from
theory~\cite{Pal:2013wfb,verley:vss,Manikandan:2017awd,Manikandan:2018erf}
-- a colloidal particle
in a stochastically shaken optical trap. This model was first experimentally studied in ~\cite{gomez:ssw}.  We study it again in order to understand the limitations posed by experimental setups for our inference scheme as well as test and benchmark our scheme for a system where the results are known.

We trap a polystyrene particle in an optical trap; further details of how the experiment
is performed may be found in the methods section.
We modulate the position of the center of the trap
$\lambda(t)$ along a fixed direction $x$ on
the trapping plane perpendicular to the beam propagation $(+z)$.
The modulation is a Gaussian Ornstein-Uhlenbeck noise with zero mean
and covariance $\bra{\lambda(0)\lambda(s)} = A \tau_0 \exp(-\mid s \mid/\tau_0)$, i.e.,
\begin{equation}
\label{slidp:OU}
\dot{\lambda}(t)=-\frac{\lambda(t)}{\tau_0}+\sqrt{2A}\eta\/,
\end{equation}
where $\eta$ is Gaussian, has zero-mean and is white-in-time. 
The correlation time $\tau_0 $
is held fixed for all our experiments. {\color{black} Note that $A \tau_0$ can be interpreted as an effective temperature \cite{dieterich:ET}}.

The  dynamics of the colloidal particle is well described by an overdamped Langevin equation,
\begin{equation}
\label{slidp:x}
\dot{x}(t)=-\frac{K}{\gamma}\left[ x(t)-\lambda(t) \right]+\sqrt{2D}\xi\/,
\end{equation}
where $K$ is the spring constant of the harmonic trap, $\gamma$ is the drag coefficient,
$\xi$ is the thermal noise, $D = \kB T/\gamma$ is the diffusion coefficient of the particle and
$T$ the temperature of the medium.
The noise $\xi$ is also Gaussian, zero-mean and white-in-time and mutually independent
from the noise $\eta$ in Eq.~(\ref{slidp:OU}).
Equations \eqref{slidp:x} and \eqref{slidp:OU} together define the model we call the Stochastic Sliding parabola.
Starting from arbitrary initial conditions for $x$ and $\lambda$, the system reaches a
non-equilibrium steady state, with the probability distribution function and current
given respectively by \cite{Pal:2013wfb}
\begin{subequations}
    \begin{align}
\label{slidp:pst}
P(x,\lambda)&=\scalebox{1}{$\frac{\exp \left(-\frac{(\delta+1) \left(\delta^2\theta (x-\lambda)^2+\delta \left(\theta x^2+\lambda^2\right)+\lambda^2\right)}{2D\tau_0\theta \left(\delta^2 (\theta+1)+2 \delta+1\right)}\right)}{2 \pi  \sqrt{\frac{D^2\tau_0^2\theta \left(\delta^2 (\theta+1)+2 \delta+1\right)}{\delta (\delta+1)^2}}}$}\/,\\
\label{slidp:pstj}
J(x,\lambda)&=\scalebox{1}{$\left(\begin{array}{c}
         \frac{\delta  \left(\delta ^2 \theta  (\lambda-x)+\delta  \lambda+\lambda\right)}{ \left(\delta ^2 (\theta +1)+2 \delta +1\right)\tau_0}  \\
          -\frac{\delta ^2 \theta (\delta  x+x-\delta  \lambda)}{ \left(\delta ^2 (\theta +1)+2 \delta +1\right)\tau_0}
    \end{array}\right)$}P(x,\lambda)\/,
    \end{align}
\end{subequations}
where the dimensionless parameters $\theta$ and $\delta$ are defined as,
\begin{align}
\delta&=\frac{K\tau_0}{\gamma},& \theta&=\frac{A}{D}.
\end{align}
The rate of entropy production and the thermodynamic force field for this model are,
\begin{subequations}
\label{eq:sigmaF}  
\begin{align}
\sigma&=\frac{\delta^2\theta}{(\delta+1)\tau_0}\/,\\
\bm{\mathcal{F}}(\xx)&\equiv
\begin{pmatrix}
  \mathcal{F}_x \\
  \mathcal{F}_\lambda 
\end{pmatrix}
= 
\left(\begin{array}{c}
         \frac{\delta  \left(\delta ^2 \theta  (\lambda-x)+\delta  \lambda+\lambda\right)}{D\tau_0 \left(\delta ^2 (\theta +1)+2 \delta +1\right)}  \\
          -\frac{\delta ^2 (\delta  x+x-\delta  \lambda)}{D\tau_0 \left(\delta ^2 (\theta +1)+2 \delta +1\right)}
    \end{array}\right)
\end{align}
\end{subequations}

In Fig. 1 we compare the above exact results to the outcome of the inference algorithm applied
to numerically generated data for this model.
Different sets of time-series data were generated by varying the noise amplitude ratio
$\theta$ by varying $A$, keeping the other parameters fixed.
 In Fig, 1b, we see that the inference algorithm predicts a value $\sigma_L$ \textcolor{black}{which is lower than the true value $\sigma$ in the beginning, but gets} very close
to the true value,  after a relatively modest number of steps. 
As we run the algorithm longer,  $\sigma_L$ saturates to something very close to the actual value.
The inference algorithm also simultaneously gives an optimal force field
${\bm d^*(x)}$ which is very similar to the thermodynamic Force
field $\bm{\mathcal{F}_{\mu}(x)}$ expected from theory (see Fig. 8 in the Methods section).
From Eq.\ \eqref{eq:sigmaF}, it is clear that $\sigma$ increases linearly with $\theta$ or equivalently the parameter $A$. Fig. 1c illustrates that the inference algorithm captures this behaviour accurately. Since we are limited by the minimal resolution of the time series in probing the $\Delta t \rightarrow 0$ limit of Eq.\ \eqref{eq:unc}, the inferred value of entropy production is in general different from the exact value by an $\mathcal{O}[\Delta t]$ term. For this model we can also compute this correction analytically as (using expressions previously obtained in \cite{Manikandan:2018erf}),
\begin{align}
\label{correctiondt}
   \sigma_{\Delta t}=\sigma -\frac{\delta ^4 \theta ^2 \left(\delta ^2 (\theta +1)+1\right)}{(\delta +1)^2 \tau_0^2 \left(\delta ^2 (\theta +1)+2 \delta +1\right)} \Delta t +\mathcal{O}[\Delta t]^2,
\end{align}
where $\sigma_{\Delta t}$ is the result one gets from Eq.\ \eqref{eq:unc} for a fixed value of $\Delta t$. 
Notice that the $\mathcal{O}[\Delta t]$ correction increases with the value of $\theta$.  The inferred values of $\sigma$ indeed lie between these two limits. 
\begin{figure}[H]
    \centering
    \includegraphics[scale=0.19]{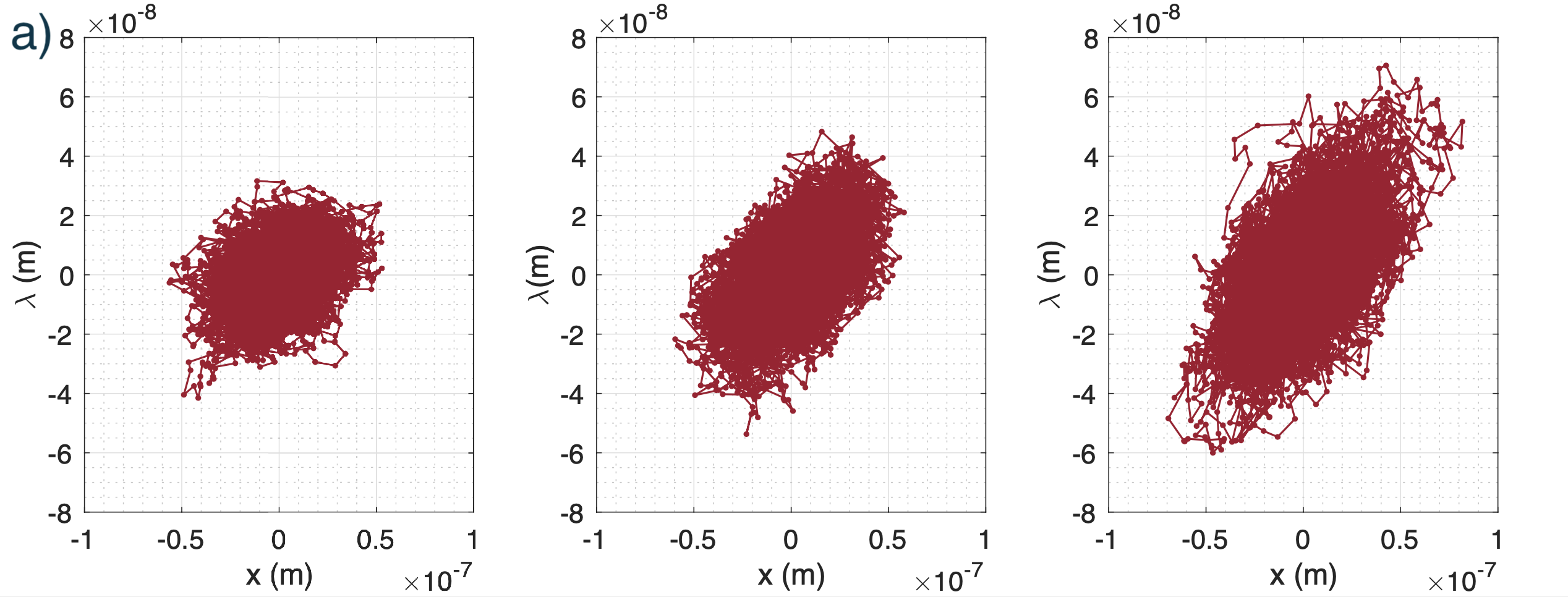}
    \includegraphics[scale=0.3]{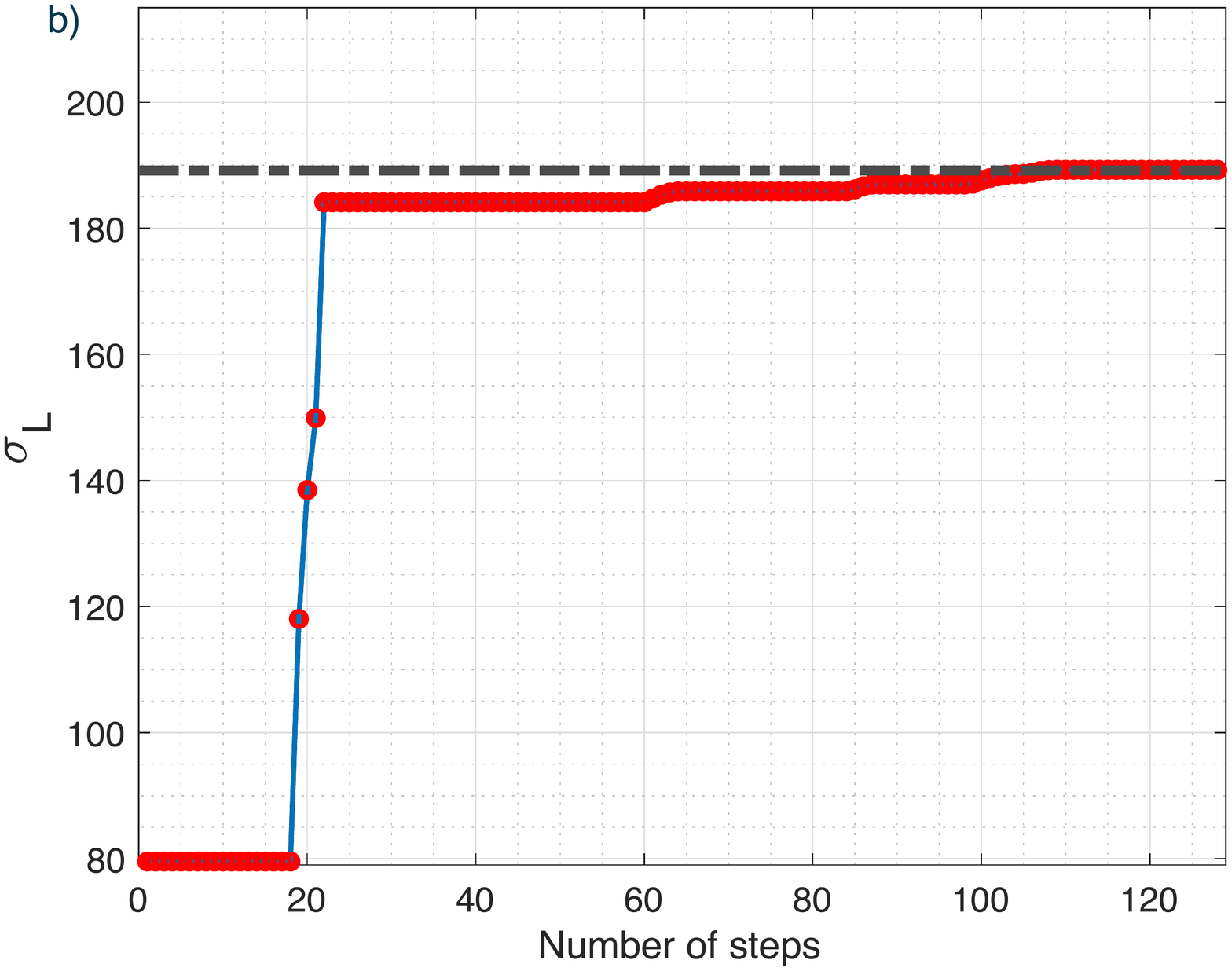}
    \includegraphics[scale=0.4]{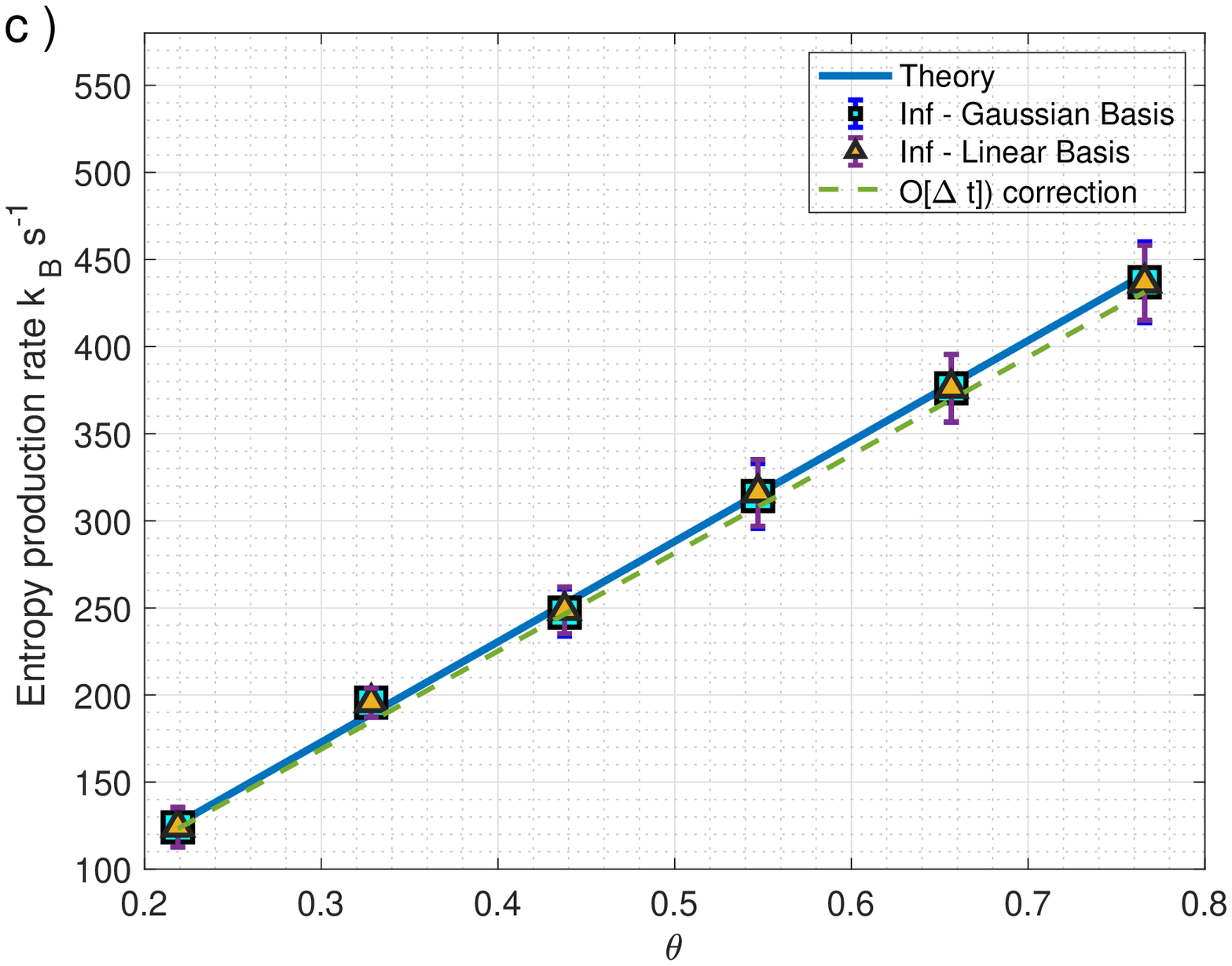}
    \caption{The inference algorithm tested on numerically generated data.  \textit{a)} Brownian trajectories of the \textit{Stochastic sliding parabola} for $\theta=0.22$, $0.55$ and $0.77$. \textit{b)} The inferred entropy production rate ($\sigma_L$) plotted against the number of steps in the optimization process for $\theta=0.33$ with $\Delta t = 0.0001$. 
    \textit{c)} Inferred entropy production as a function of the parameter $\theta$. The blue line corresponds to the theoretical value of $\sigma$, Eq. \eqref{eq:sigmaF}. The squares correspond to the inferred values of $\sigma$ using the inference scheme (Eq.\ \eqref{eq:unc}) in the Gaussian basis, and triangles correspond to inference using the Linear basis. The green dashed line corresponds to the best estimate of $\sigma$ that can be obtained using inference with $\Delta t=0.0001$,  using Eq.\ \eqref{correctiondt}. The error estimates are set by computing the standard deviation over the values obtained for an ensemble of $8$ trajectories ( see Fig. $7$ in the Methods section).}
    \label{fig:my_label}

\end{figure}

Next, we tested the algorithm on experimentally generated data for the same model. In the experiments, we varied $A$ ranging from $0.1$ to $0.35$ {in units of} 
$\left(0.6 \times 10^{-6}\right)^2\;{\rm m}^2{\rm s}^{-1}$ (corresponds to $\theta$ varying from $0.22$ to $0.77$), while the other experimental parameters such as the trap stiffness, as well as the bath temperature, were assumed to be constant for the entire length of the experiment. In reality however, the laser used to trap the particle is prone to power fluctuations, and there can also be minor changes in the bath temperature due to  heating caused by the long exposure to the laser. \textcolor{black}{For large values of $\theta$, we also expect  non-harmonic effects to be significant, due to the particle exploring the peripheral regions of the trap \cite{richardson2008non}}.  We comment in the following paragraph on the implications of these fluctuations for our results. An immediate consequence is however that the theoretically predicted values  Eq. (12a) can only be used as a reference. We benchmark  our results instead by comparing them with values obtained by the application of the Stochastic Force Inference technique (SFI) scheme recently proposed in \cite{frishman:sfi}, which gives an independent estimate of both $\sigma$ as well as the force fields.

Experiments for individual parameter sets were carried out for a duration of 100s, with a sampling rate of 10 KHz for the particle position. Only about $2/3$rd of the available experimental data was used and the remaining
$1/3$rd was discarded because of the presence of uncontrolled experimental errors in them. For the analysed data, each of the 100s long data sets were further divided into 12.5s long patches, upon which  the inference algorithm was then tested. In Fig.~2, we demonstrate the results of the analysis of the experimental data. The dark-blue dashed  line corresponds to the theoretically predicted value,  Eq.\ \eqref{eq:sigmaF}, of the entropy production rate  for the model given by Equations \eqref{slidp:OU} and \eqref{slidp:x} with the given parameters (the values are given in the Supplemental information). The blue line is the entropy production for a slightly modified model explained in the Supplemental information, obtained by analysing the data obtained from SFI and calculating the drift and diffusion terms from it. The region between the red dashed lines corresponds to the error bar set by the variation in model parameters (namely the drift and diffusion coefficients) in different experiments as quantified by the SFI analysis (see the Supplemental Information).  The data points are the results of our inference algorithm as well as the SFI scheme. As is evident, our inference scheme predicts exactly the same or very similar values for $\sigma$ as the SFI algorithm, for all values of $A$. 

The prediction of the inference scheme and SFI matches also for the thermodynamic force (see Fig. \ref{fig:OptF2}). Namely, the optimal current $\textbf{d}^*(\textbf{x})$ which we get as an outcome of our inference algorithm, also matches 
${\hat{\mathcal{F}}(\textbf{x})}$, which is ${\mathcal{F}(\textbf{x})}$ estimated from the trajectory data by means of the Stochastic Force Inference technique \cite{frishman:sfi}. We conclude that our inference
algorithm infers the correct entropy production value, as well as the correct thermodynamic force field, for the experimental data, since we get the same results when using a completely independent and different technique. 
After this benchmarking exercise,  we study now a more complicated situation.
\begin{figure}
    \centering
    \includegraphics[scale=0.4]{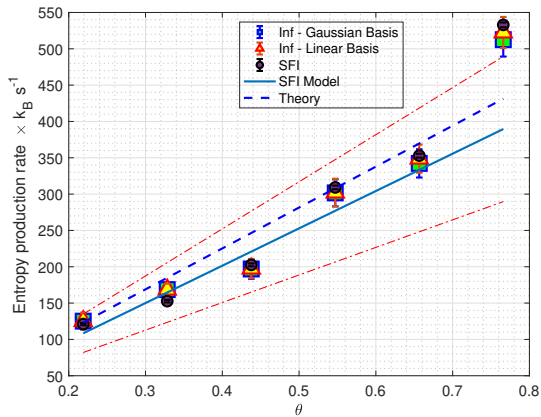}
    \caption{ Our inference algorithm tested on the experimental data for different values of the parameter $A$ (or $\theta $). The dark-blue dashed line corresponds to the theoretical value given by Eq.\ \eqref{eq:sigmaF}. The squares and triangles corresponds to $\sigma$ estimated from the experimental data using our TUR-based inference scheme with a Gaussian basis and a Linear basis, and using $\Delta t=0.1ms$. The error bars correspond to averages over eight independent realizations of duration $12.5s$. The circles correspond to $\sigma$ estimated using the Stochastic force inference scheme (SFI) \cite{frishman:sfi} for the whole $100s$ data set, and the errorbars for these correspond to a self-consistent estimate of the inference error that the SFI provides \cite{frishman:sfi}. The blue line corresponds to $\sigma$ predicted by a model obtained from SFI (see Supplemental Info), and the red dashed lines correspond to error bars for this SFI-based model.   }
    \label{fig:exptt}
  \end{figure}

\begin{figure}
    \centering
\includegraphics[scale=0.25]{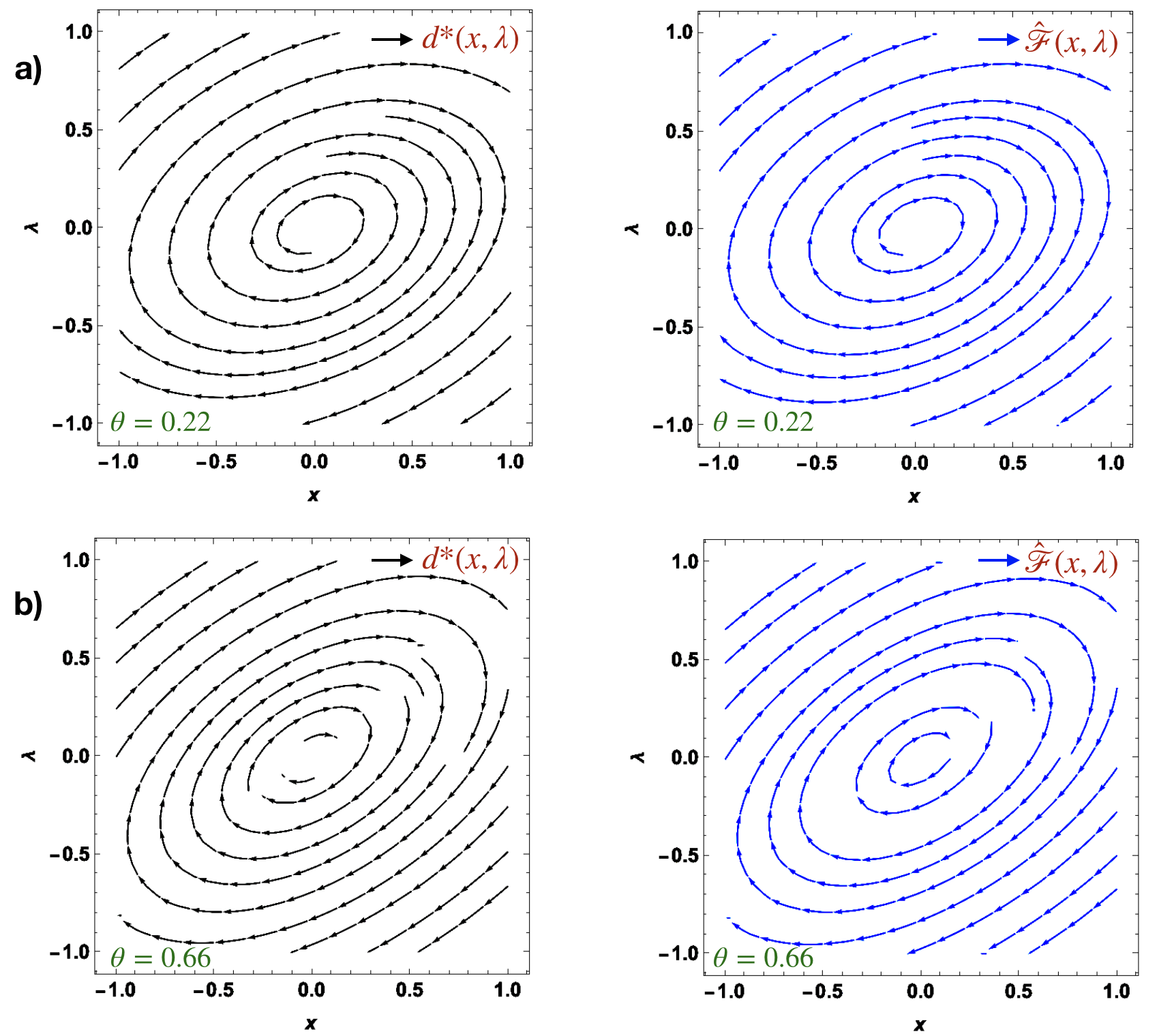}
 \caption{ Optimal force fields (streamline plots) obtained from the experimental data: $\textbf{d}^*(x,\lambda)$ (black) compared to $\bm{\hat{\mathcal{F}}}(x,\lambda)$, which is $\bm{\mathcal{F}}(x,\lambda)$ estimated from the data using the Stochastic Force inference technique \cite{frishman:sfi}, (blue) in two cases. The parameter choices used are a) $\theta =0.22$ and b) $\theta=0.66$. We see that $\textbf{d}^*(x,\lambda)$ agrees well with $\bm{\hat{\mathcal{F}}}(x,\lambda)$.    }
    \label{fig:OptF2}
\end{figure}

\subsubsection{A colloidal particle trapped near a microbubble}
\begin{figure}
    \centering
    \includegraphics[scale=0.25,angle=-90]{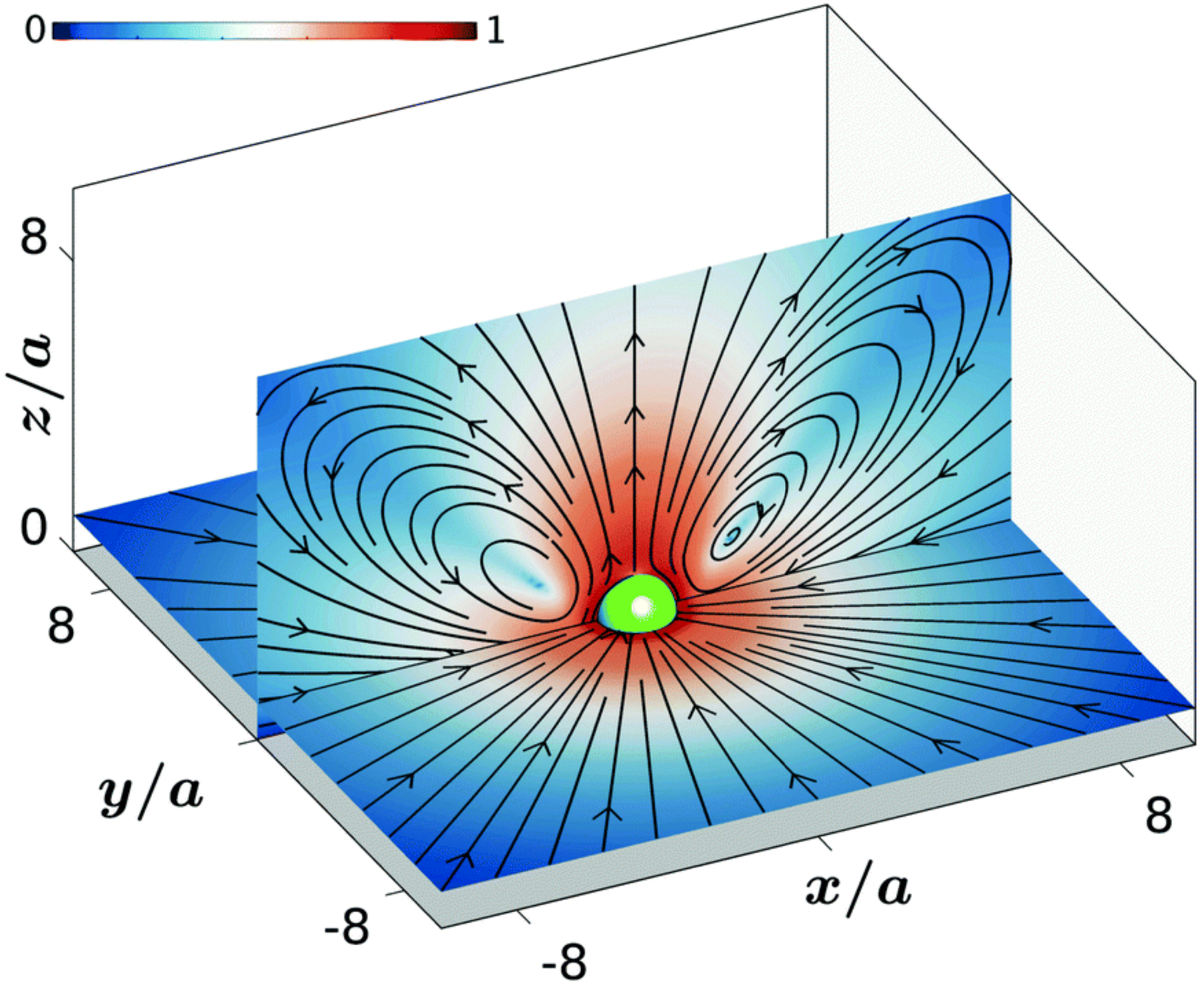}
    \caption{ Cross-section of the streamlines of fluid flow around a bubble in planes parallel and perpendicular to the wall (shown in grey at $z = 0$). It can be seen that the flow has cylindrical symmetry and draws fluid from all directions. The streamlines of the fluid flow are drawn over the pseudo-color plot of the normalized logarithm of the flow speed. Figure and caption taken from \cite{ghosh2019self} - Reproduced by permission of The Royal Society of Chemistry}
    \label{fig:flow}
\end{figure}

\begin{figure}
    \centering
    \includegraphics[scale=0.16]{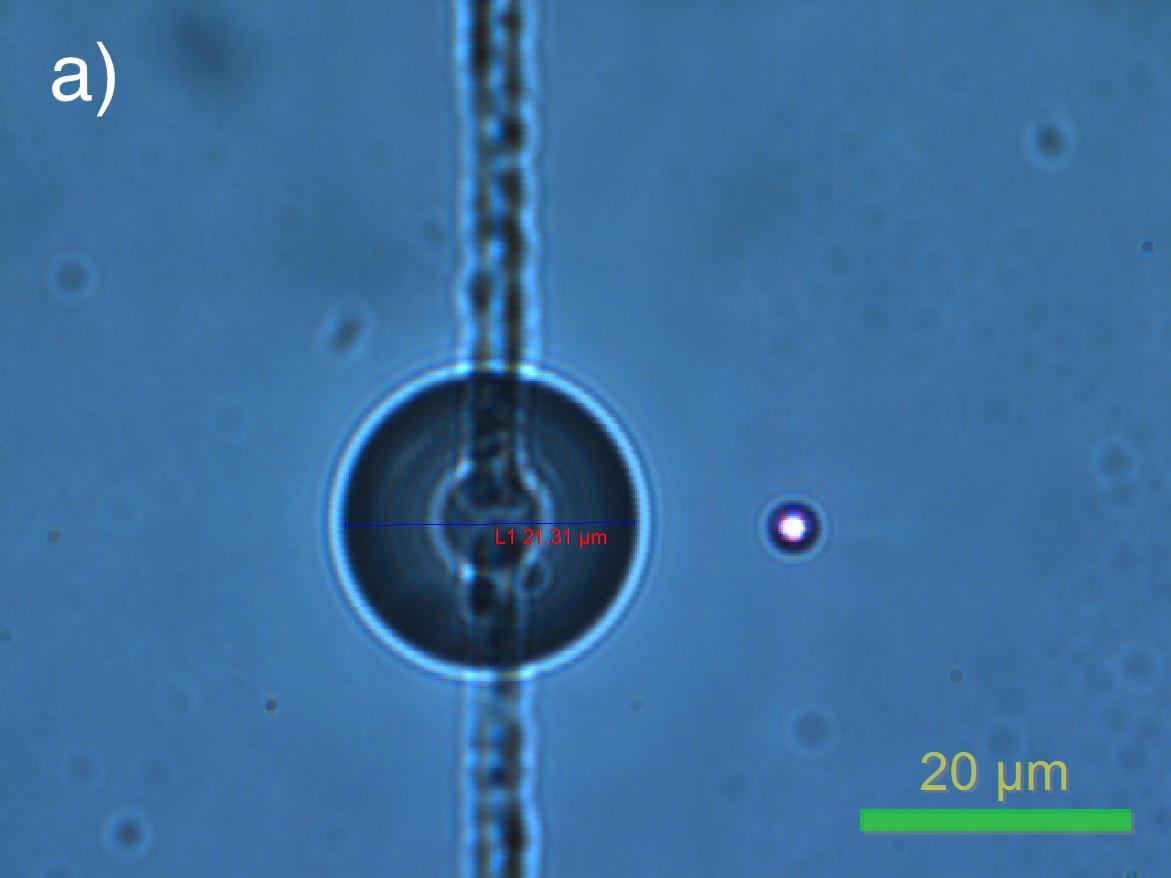} \\
      \includegraphics[scale=0.28]{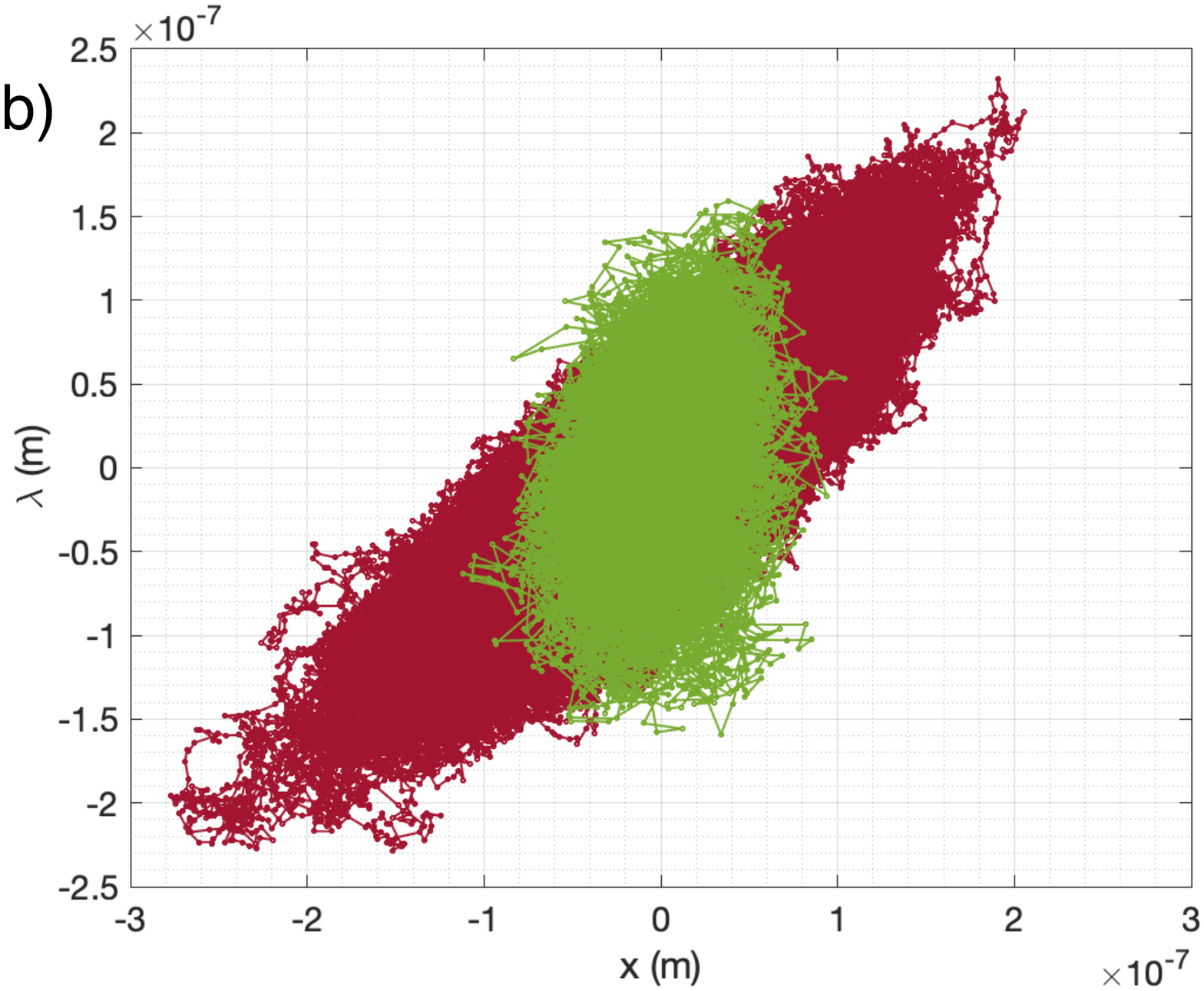}\\
     \includegraphics[scale=0.345]{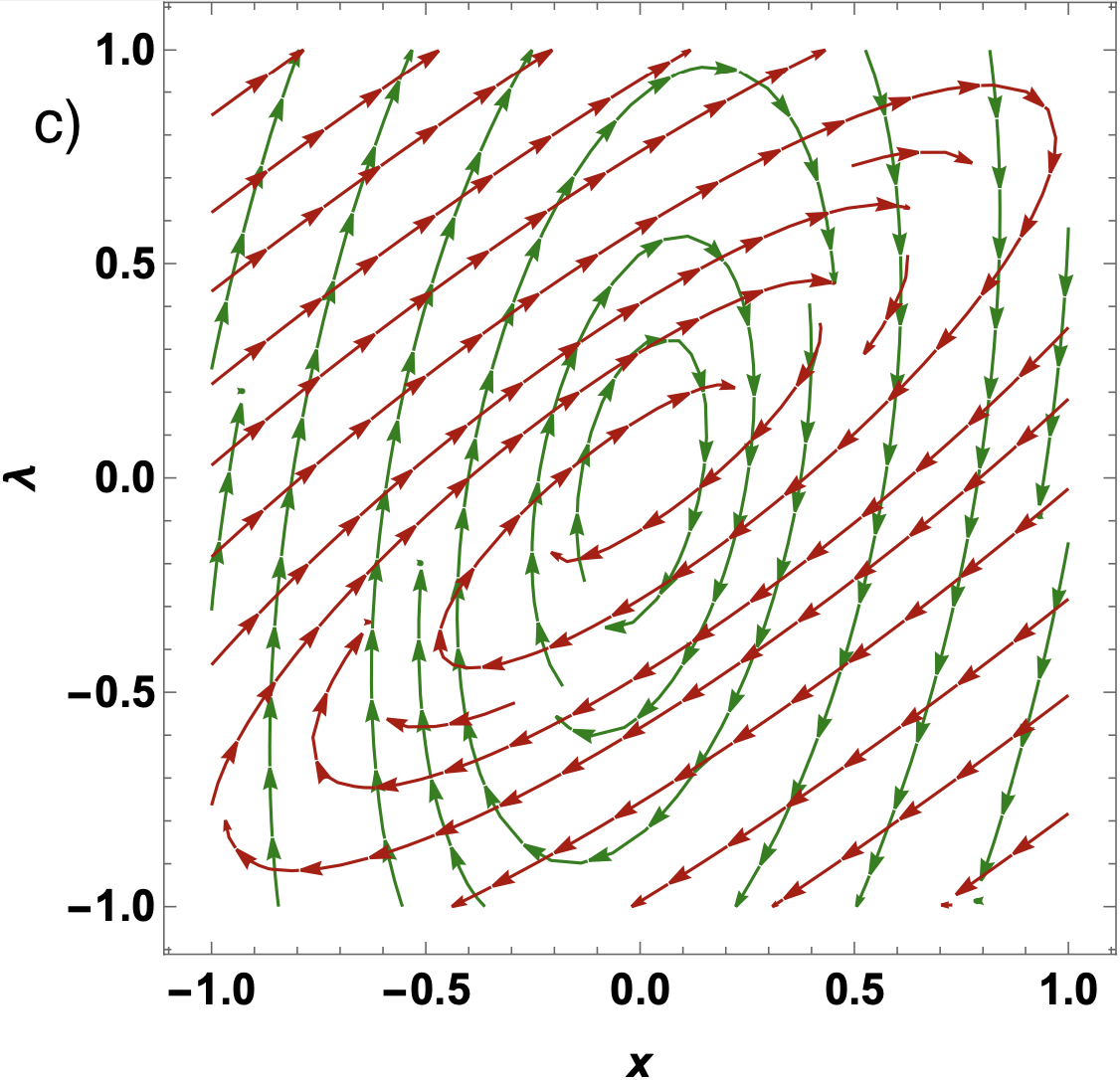}\;\;\;\;\;\;\;\;
    \caption{The colloidal system in the presence of the bubble. a) The microbubble - colloidal particle system. b)  System trajectories without (red) and with (green) the bubble in the neighbourhood of the colloidal particle. We see that the colloidal particle is strongly confined in the presence of the bubble. c) The thermodynamic force field computed as the optimal field $\textbf{d}^*(\textbf{x})$ without the bubble (red) and in the presence of the bubble (green). The corresponding entropy production rates estimated are $\sigma = 244.68\; k_B s^{-1}$ for the no-bubble case and $\sigma = 7.66\; k_B s^{-1}$ for the case with the bubble.  } 
    
    \label{fig:bub1}
\end{figure}

 We now show how our scheme performs in estimating the rate of entropy generation for the
case where the mechanical force on the colloidal particle is not known.
For this purpose, we study a  particle trapped in the vicinity of a microscopic bubble
of size 20-22 $\mu m$.
We have already used this experimental setup to study one or more microbubbles with  colloidal
particles moving in the liquid in a different context~\cite{roy2016exploring,ghosh2019self}.
The microbubbles are nucleated on a liquid-glass interface.
The surface is pre-coated by linear patterns of a MB-based soft oxometalate (SOM)
material.
We focus a laser beam on any region along this pattern, the SOM material gets intensely
heated and a microbubble forms.
The top of the bubble is colder than its bottom where it is anchored to the interface. 
As the surface tension is a function of temperature, the variation of the surface
tension along the surface of the bubble sets up a Marangoni stress, driving a flow
along the surface of the bubble.
Marangoni flow around freely floating bubbles under a temperature gradient have been studied
both experimentally~\cite{hardy1979motion} and analytically~\cite{young1959motion}.
The additional complexity here is the presence of the bottom surface on which the flow
must satisfy no-slip boundary conditions. 
The flow around the bubble in this setup is not yet known in detail although
an approximate description, valid if we are not too close to the bubble,
has been developed~\cite{ghosh2019self}, which we show in Fig.~\ref{fig:flow}.
This flow drags  the trapped colloidal particle and changes its steady-state probability
distribution (See Figs.\ \ref{fig:bub1}a and b). Since the flow streamlines are directed towards the bubble, we expect that these will confine the trapped particle more than the case without the bubble. This is indeed the case as we show later.
 We expect that the underlying description of the particle is still an overdamped Langevin equation, including a flow velocity field $u(x)$. However, the quantification of this flow field is rather difficult, even numerically, as argued above. As a result, we have a system where the details of the microscopic description and forces are unknown. Our inference scheme, on the other hand, is easily applicable even in this context. 

\begin{figure}
    \centering
     \includegraphics[scale=0.4]{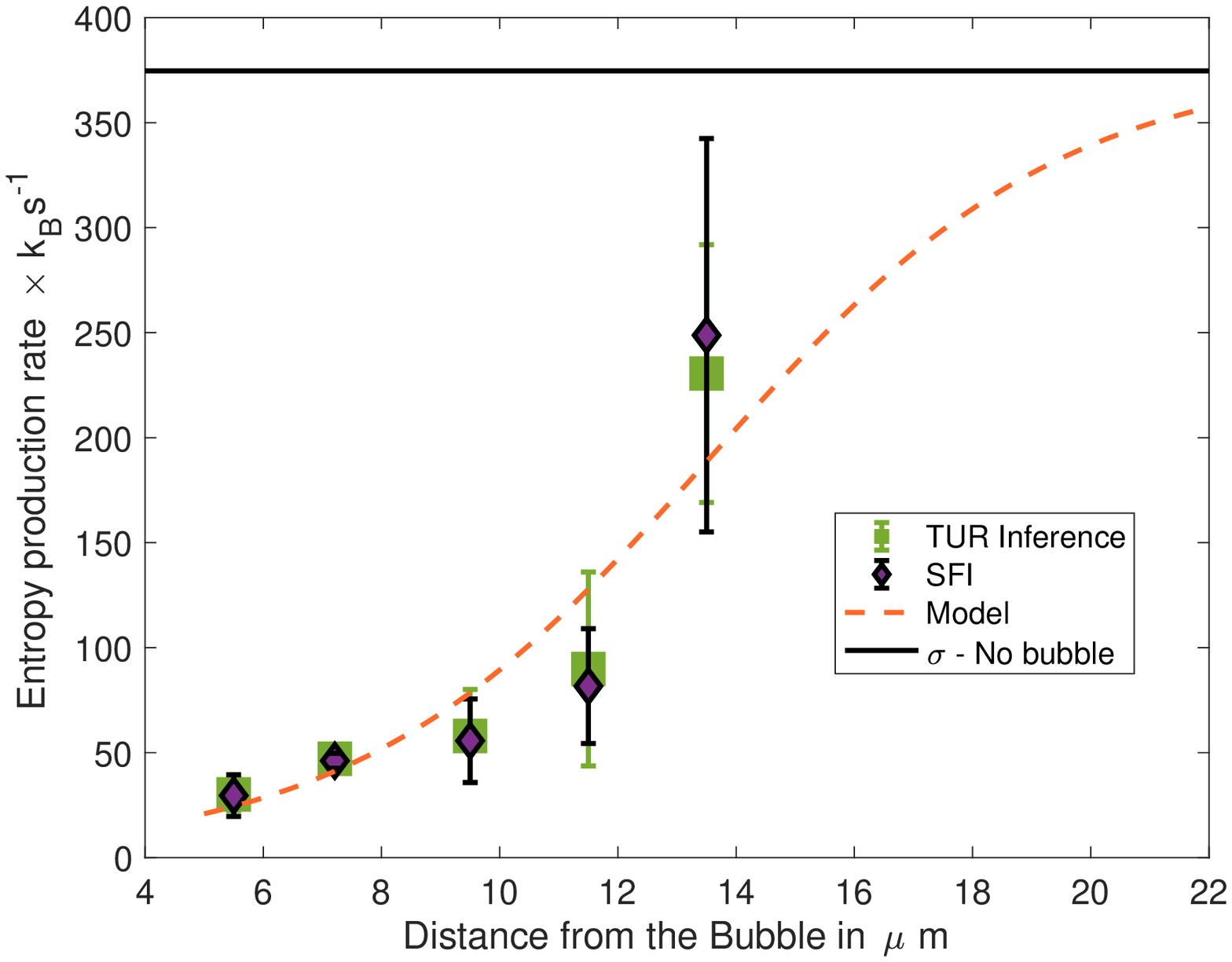}
    \caption{Short-time TUR and SFI estimates for the entropy production rate in the colloidal system in the presence of the bubble, as a function of the distance from the surface of the bubble. The inferred value of $\sigma$ agree for the two different algorithms as well as with the value predicted by the ad-hoc model in Eq.\ \eqref{eq:bubblemodel}. The fit parameters used for the free parameters in the model are, $a=282.743$ and $b=1/3 \mu m^{-1}$. The error estimates are standard deviations over an ensemble of $6$ trajectories, each of  length $10 s$ and $\Delta t = 0.0001 s$, for both the  schemes.  }
    \label{fig:bub2}
 \end{figure}

At the level of the non-equilibrium trajectories of the system, we see that there is a qualitative difference from the case without the bubble. First, we see that the particle is more confined in the trap along the $x$-direction, when there is a bubble in the vicinity (see Fig.\ \ref{fig:bub1}b) as mentioned earlier. This confinement is  caused both by the flow towards the bubble (as shown in Fig. $4$), which gets balanced at the confined position by the opposing force of the confining potential, as well as the reduced fluctuations close to the bubble due to proximity effects \cite{Hydromob}. Further statistical analyses also reveal weaker non-equilibrium currents (see Fig. 14 in the Supplemental Information).
Consistent with these observations, on applying the inference algorithm, we observe that 
the value
of $\sigma$ is substantially reduced in the presence of the bubble. The corresponding entropy production rates estimated are $\sigma = 244.68\; k_B s^{-1}$ for the no-bubble case and $\sigma = 7.66\; k_B s^{-1}$ for the case with the bubble. We also find that the thermodynamic force, estimated using the inference scheme, is significantly reduced along the x direction, and the force field is less tilted along that direction as compared to the case without the bubble, as shown in Fig.\ \ref{fig:bub1}c. 

To further analyze the effect of the bubble, we performed another experiment, where we trapped the particles at different distances from the bubble. As we go  a distance $d \sim 1.5r$ ($r$ is the radius of the microbubble) from the surface of the bubble, we see that the inferred value of $\sigma$ gets closer to the value the system would have had in the absence of the bubble. This is demonstrated in Fig.\ \ref{fig:bub2}.

An  important point to understand here, in the light of these findings is the significance of the inferred value of $\sigma$. In the case without the bubble, it is exactly the total heat dissipated to the environment as a consequence of maintaining the system in a non-equilibrium steady state (by shaking the trap). In the case with the bubble however this is not the case. We present a possible mathematical description of this situation as an overdamped Langevin equation  with space-dependent diffusion and damping terms in an unknown flow field $u(x)$. Since the trap constrains the particle motion on scales which are at least two orders of magnitude smaller than the distance to the bubble, $u(x)$ is further assumed to be a constant $u_d$ at a distance $d$ from the surface of the bubble. $\sigma$ calculated from this model, reproduces the values we find from the experimental data, independent of $u_d$, and purely as a consequence of the space-dependent diffusion and damping term, and the two fitting parameters $a$ and $b$. This is demonstrated in Fig.\ \ref{fig:bub2} .  As we discuss in the supplemental material however, there is another component of the entropy production, related to the work that the flow does against the confining potential \cite{Seifert:ref,andrieux2008thermodynamic}. This component, which does indeed depend on the value of $u_d$, is not estimated by our inference scheme, due to the fact that $u_d$ is a field (corresponding to the velocities of the molecules of the thermal bath) which is odd under time reversal, for which the TUR does not hold \cite{Seifert:inf,TURunderdamped1,TURunderdamped2,fischer2020free,niggemann2020field}. Hence we expect that the values of $\sigma$ we find close to the bubble are underestimates of the true value. We elaborate on this point in the supplemental information.

\paragraph{Mathematical model: } The colloidal system in the presence of the bubble and consequently the flow $u_d$, can be simulated using the following equations:
\begin{align}
\label{eq:bubblemodel}
\begin{split}
    \dot{x}-u_d &= -\frac{\left(x-\lambda \right)}{\tau_d}+\sqrt{2D_d}\;\eta(t),\\
    \dot{\lambda}&=-\frac{\lambda}{\tau_0}+\sqrt{2A}\;\xi (t),
    \end{split}
\end{align}
where,
\begin{align}
\begin{split}
    \tau_d&=\tau \left(a \exp(- b d )+1\right),\\
    D_d&=\frac{D}{a \exp(- b d )+1}.
       \end{split}
\end{align}
 Here the parameters $a$ and $b$  can be tuned to match the experimental data. Particularly, $1/b$ stands for a characteristic length scale below which the flows created by the bubble are significant. When the distance of the trapped particle from the bubble is  much greater than $1/b$, we expect that the expressions will match  the case without the bubble. Using a trial and error approach we obtained the fit parameters as $a=282.743$ and $b=1/3 \mu m^{-1}$. 

We remark that, as we did for the case without the bubble, the SFI technique could be used to model this case as well, since it explicitly gives the drift and diffusion terms. These are however particularly susceptible to erroneous estimates of a {\it conversion factor} which is needed to obtain the particle trajectory data in units of \textit{nm}. We expand on this issue in the methods section as well as in the supplemental material.  This error can be thought of as assigning wrong units to the affected phase-space coordinates. Since $\sigma$ is a sum over all phase space coordinates, its evaluation is not affected by such an error unlike other quantities such as forces, diffusion terms and the thermodynamic force. Another way to understand this is to note that $\sigma$ quantifies the irreversibility of the dynamics, which again is clearly not affected by a choice of units. Hence our model for the setup  with the bubble only tries to reproduce the value of $\sigma$ as a function of distance.

 In conclusion, we have experimentally tested a simple and effective method, based on the Thermodynamic Uncertainty Relation \cite{manikandan2019inferring,Shun:eem,van:epe} for inferring both the rate of entropy production $\sigma$ and the corresponding thermodynamic force fields, in microscopic systems in non-equilibrium steady states. We have confirmed that an entirely independent method, SFI \cite{frishman:sfi}, gives the same answers in all the situations we have studied, hence adding weight to the physical significance of our findings.  We have also carried out an extensive investigation of the convergence properties of our code as several parameters or hyper-parameters are varied, as well as a comparison with the SFI algorithm (Figs $9-13$) in the Methods section. 

 Our short-time inference scheme does not need any model in order to be applicable. However, we can use our findings to come up with plausible models, which give the same $\sigma$ values for a range of parameters, even in cases where modeling the system from first principles is complicated. 
 In this regard it would also be interesting to perform a systematic study of different algorithmic schemes available to model a complex non-equilibrium systems, with focus on the advantages and disadvantages when applied to experimental data.


 
 Experimental systems that would be particularly interesting to study are molecular motors or other cellular processes.
 Recently, Ref.~\cite{hurst2021intracellular} tried to quantify the activity of a cell by measuring the power spectral density
 of the fluctuations of the position of a phagocytosed micron-sized bead {\em inside a cell}.
 As it is possible to also trap such beads inside a cell with optical tweezers~\cite{hurst2021intracellular},
 this too could be a very interesting system to study.
 Finally, in other recent work \cite{otsubo2020estimating}, it has been demonstrated that inference schemes of this kind can also
 be made to work for non-stationary non-equilibrium states, further diversifying the scope of this class of techniques.



\section*{Acknowledgement}
DM and VA acknowledge the support of the Swedish Research Council through grants 638-2013-9243 and
2016-05225. SK and SKM thank Shun Otsubo for helpful discussions. SKM thanks Ralf Eichhorn for pointing out a useful reference.
\section*{Author contribution Statement}
SK, SKM, AB, DM and SG designed research; SG, AK and BD performed the experiments in AB's lab; VA, BD, SKM and DM implemented the algorithm; VA, BD, SKM and SG analyzed the data; all the authors discussed the results; SKM, SG, SK, DM and AB together wrote the manuscript. SKM and SG contributed equally to the work.
\section*{Code Availability}
We use the open-sourced \textit{PYSWARM} package in Python \cite{pyswarms} for the optimization task. The algorithm used to produce the results in this paper is available at: https://doi.org/10.6084/m9.figshare.14174369. We perform the Stochastic Force Inference (SFI) analysis using the algorithm provided with \cite{frishman:sfi}. 

\section*{Data Availability}
The data used to produce the results in this paper is available at: https://doi.org/10.6084/m9.figshare.14176664

\section*{Competing interests }
The authors declare that they have no competing interests.
\appendix
\section{Materials and methods}
\label{app:method}
\subsection{Experiment}
\label{app:exp}
\subsubsection{A single colloidal particle in a stochastically shaken trap}
\label{s:expt}
The experimental setup consists of a sample chamber placed on a motorized xyz-scanning microscope stage, which contains an aqueous dispersion of spherical polystyrene particles (Sigma-Aldrich) of radius $r = 1.5~\mu m$. The sample chamber consists of two standard glass cover-slips (of refractive index $\sim 1.52$) on top of one another. The thickness of the chamber is kept $\sim 100 ~\mu m$ by applying double-sided sticky tape in between the cover-slips. The aqueous immersion is made out of double distilled water at room temperature, which acts as a thermal bath. A single polystyrene particle is confined by an optical trap, which is created by tightly focusing a Gaussian laser beam of wavelength $1064~ nm$ by means of a
high-numerical-aperture oil-immersion objective (100x, NA = 1.3) in a standard inverted microscope (Olympus IX71).
The trap is kept fixed at a height, $h= 12~ \mu m$ from the lower surface of the chamber in order to avoid spatial variation in the viscous drag due to the presence of the wall. The corner frequency of the trap ($f_c$) is set to be $135 Hz$. For the first set of experiments, the center of the trap is modulated ($\lambda(t)$) using an acousto-optic deflector, along a fixed direction $x$ in the trapping plane, perpendicular to the beam propagation $(+z)$. Thus, the modulation may be represented as a Gaussian Ornstein-Uhlenbeck noise with zero mean and
covariance $\bra{\lambda(s)\lambda(t)} = A \tau_0\exp(\vert t-s\vert/\tau_0)$.
The correlation time $\tau_0 $ is held fixed for all our experiments. We determine the  barycenter $(x, y)$ displacement of the trapped particle by recording its back-scattered intensity from a detection laser (wavelength 785 nm, co-propagated with the trapping beam) in the back-focal plane interferometry configuration. The measurement is carried out using a balanced-detection system comprising of high-speed
photo-diodes \cite{bera2017fast}, with sampling rate of $10$ kHz and final spatial resolution of~$10~$nm. In all cases, the trap parameters including the conversion factors for the trajectory data were calibrated by fitting the probability distribution of the particle position in
thermal equilibrium to the Boltzmann distribution $P(x)=(2\pi D\tau )^{-1/2} \exp(- x^2/2D\tau)$, where $\tau$ is the relaxation time in the trap, given by $\tau = 1/(2\pi f_c)$. We assume that the diffusion constant $D$ has the room temperature value, $D=1.645\times 10^{-13} m^2s^{-1}$.  It is important to note that we assume that the trap parameters as well as the conversion factor for the trajectory data are unaffected  when the Ornstein-Uhlenbeck modulation is turned on. However, in practice, the trap parameters can indeed be altered by small amounts over long durations of measurement ( - 100s ), primarily due to the power fluctuations of the trapping laser. Further, the probe particle also moves in the $y$ and $z$ directions in the trap, which we have not measured here. These factors led to issues which prevented us from producing an exact replica of the theoretical model in the experiment. However, we have taken into account these limitations in our analysis as detailed in the supplemental information, Section I.

In the second set of experiments, i.e. for those with the microbubble, we employ a cover slip that is pre-coated by a polyoxometalate material  \cite{ghosh2018assembly,ghosh2021review} absorbing at 1064 nm as one of the surfaces of the sample chamber (typically bottom surface), and proceed to focus a second 1064 nm laser on the absorbing region. A microbubble is thus nucleated - the size of which is controlled by the power of the 1064 nm laser \cite{ghosh2018assembly}. Typically we employ bubbles of size between 20-22 $\mu$m. Note that the sample chamber also contains the aqueous immersion of polystyrene particles. We trap a polystyrene probe particle at different distances from the bubble surface, and modulate the trap centre in a manner similar to the experiments without the bubble. The particle is trapped at a axial height corresponding to the bubble radius. The other experimental procedures remain identical to the first set of experiments. An important point here though, is the determination of the distance of the particle from the bubble surface. This we accomplish by using the pixels-to-distance calibration provided in the image acquisition software for the camera attached to the microscope, which we verify by measuring the diameters of the polystyrene particles in the dispersion (the standard deviation of which is around 3\% as specified by the manufacturer), and achieve very good consistency. Note that we obtain a 2-d cross-section of the bubble as is demonstrated in Fig.~\ref{fig:bub1}, and are thus able to determine the surface-surface separation between the bubble and the particle with accuracy of around 5\%. During the experiment, we also ensure that the bubble diameter remains constant by adjusting the power of the nucleating laser - indeed the bubble diameter is seen to remain almost constant for the 100 s that we need to collect data for one run of the experiment.

As opposed to the previous setting, the particle is now trapped in a region where it experiences: 1) a temperature gradient created by the laser beam used to generate the micro-bubble, 2) the microscopic flow generated due to the bubble which affects particle trajectory, and 3) Faxen-like corrections to the viscous drag coefficient of the encompassing fluid due to the proximity to a wall \cite{Brenner} - which is the bubble surface in this case. Now, since the trap parameters - particularly the conversion factor for the trajectory data - are determined for the equilibrium setting, it is clear that there could be significant deviations from those in the non-equilibrium configuration produced due to  the presence of the bubble in close proximity of the trapped probe particle. It is also clear that the previous procedure for obtaining the trap parameters will not work in this case. This severely hampers any exact modeling and as a result we concentrate on getting only the value 
of $\sigma$ and its variation as a function of the distance to the bubble, both of which are robust against the above errors.

\subsection{Numerical algorithm}
\label{app:num}
Our aim is to maximise a cost function $\mathcal{C}$ which is a function of a set of parameters $\bm{w}$.
We use a  particle swarm optimization algorithm~\cite{zhang2015comprehensive} to achieve this. A domain is chosen and $N_{\rm p}$ particles are initialised in that domain. The $k$-th particle follows Newtonian dynamics given by:
\begin{subequations}
\begin{align}
 \frac{d}{dt}\bm{\omega}^{\rm k} &= \bm{V}^{\rm k} \\
 \frac{d}{dt}\bm{V}^{\rm k} &= \bm{A}^{\rm k}(\bm{\omega}) \/.
 \end{align}
 \label{eq:swarm}
 \end{subequations}
 Here $\bm{\omega}^{\rm k}$ and $\bm{V}^{\rm k}$ are the position and velocity vector of the $k$-th particle and $\bm{A}^{\rm k}$ is a stochastic function that depends on the position of \textit{all the particles}. Different variants of this algorithm use different $\bm{A}$. The simplest -- the one that we use -- is called the \textit{Original PSO}. Let us first define the following:
 \begin{itemize}
\item  The $k$-th particle carries an additional vector $\bm{P}^{\rm k}$ which is equal to $\bm{\omega}^{\rm k}$ for which the value of the function $\mathcal{C}$  as observed by the $k$-th particle was maximum in its history. 
\item  At any point of time let $\bm{G}$ denote the position of the particle in the whole swarm for which the function has the maximum value.
\end{itemize}
The function $\bm{A}$ is given by
\begin{equation}
 A^{\rm k}_{\mu} = W_{\rm 1}\delta_{\mu\nu}U^{\rm 1}_{\nu}(P^{\rm k}_{\nu} - \omega^{\rm k}_{\nu}) +  W_{\rm 2}\delta_{\mu\nu}U^{\rm 2}_{\nu}(G_{\nu} - \omega^{\rm k}_{\nu}) 
 \label{eq:Fswarm}
\end{equation}
Here the Greek indices run over the dimension of space. $W_{\rm 1}$ and $W_{\rm 2}$ are two weights. The two terms in Eq.\ \eqref{eq:Fswarm} push the particle in two different directions: one towards the point in history where the particle found the function to be a maxima and the other towards the point where the swarm finds the maximum value of the function at this point of time. These are multiplied by two random vectors $U^{\rm 1}$ and $U^{\rm 2}$ of dimension same as the dimension of space. Each of the components are independent, uniformly distributed (between zero and unity), random numbers.

We keep track of the highest value of the function seen by the swarm and also the location of that point. There are two major advantages to this  over standard gradient ascent algorithms: one, it does not require evaluation of the gradient of the function and two, it can be parallellized straightforwardly. All the numerical results reported in this paper are obtained using this algorithm. We implement this optimization scheme using open-sourced \textit{PYSWARM} package in Python \cite{pyswarms}, with a default choice for the hyper-parameters. 


\subsection{Implementation of the algorithm}
Here we describe how we applied this algorithm to numerical/ experimental data. We generate numerical data using first order Euler integration of Eq.\ \eqref{slidp:OU} and Eq.\ \eqref{slidp:x} with a time step of $\Delta t = 0.0001$.  In either case we generate many copies of trajectories of length 12.5s, and construct the cost function in Eq.\ \eqref{eq:unc} using Eq.\ \eqref{eq:JJ} and Eq.\ \eqref{eq:psi}. We have tried out two different choices of basis functions to construct ${\bm d}(\XX)$. The first one is a Gaussian basis  in which we represent ${\bm d}(\XX)$ as,
\begin{align}
     {\bm d}(\XX)=\sum_{m=1}^M {\bm \omega}_m e^{-\frac{\left(x-x_m\right)^2}{2 b_x^2}}e^{-\frac{\left(\lambda-\lambda_m\right)^2}{2b_\lambda^2}}.
\end{align}
Making use of the spatial symmetry of the problem, we assume ${\bm d}(\XX)$ to be an anti-symmetric function, with ${\bm d}(-\XX)=-{\bm d}(\XX)$, and that reduces the dimensionality of the problem by a factor of 2. Here $M$ is the number of Gaussian functions, and $b_i$ are the variance of the Gaussian in the $x$ and $\lambda$ direction. The centers of the Gaussian $(x_m,\lambda_m)$ are placed equally spaced in a rectangular region enclosing the data. Both $M$ and $b_i$ are hyper parameters. We used $M=16$ and $b_{x/\lambda}^2 = \lbrace x/\lambda\rbrace_{max}/30$. Secondly, we have also tried a linear basis (motivated by the prior knowledge of the linearity of the system) where we take
\begin{align}
     {\bm d}(\XX)= {\bm \omega}_1 x + {\bm \omega}_2\lambda.
\end{align}
 In Figs.\ \ref{fig:shortlimit}, \ref{fig:sizeofdata}, \ref{fig:Np}, and \ref{fig:Ng}, we study the dependence of the output of the algorithm, for both basis functions, on $\Delta t$, length of the time series data, $N_p$ as well as $M$. 

\begin{figure}[H]
    \centering
    \includegraphics[scale=0.5]{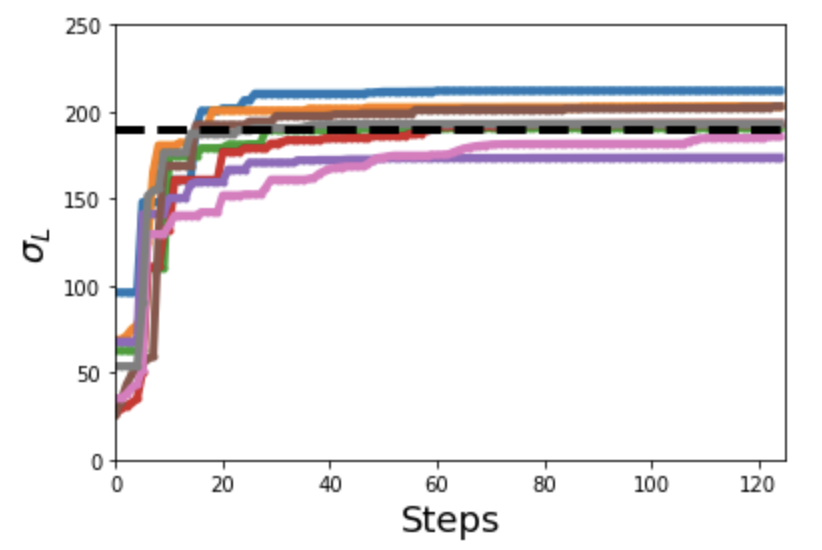}
    \caption{The value of sigma inferred ($\sigma_L$) using the Gaussian basis, as a function of the number of steps in the optimization algorithm for different 12.5s data sets, that are numerically generated for the same parameter choice as in Figure 1b of the main text. The black dashed-line corresponds to the theoretical estimate of $\sigma$ for this parameter choice.}
    \label{fig:12p5F}
\end{figure}

\begin{figure}[H]
    \centering
    \includegraphics[scale=0.4]{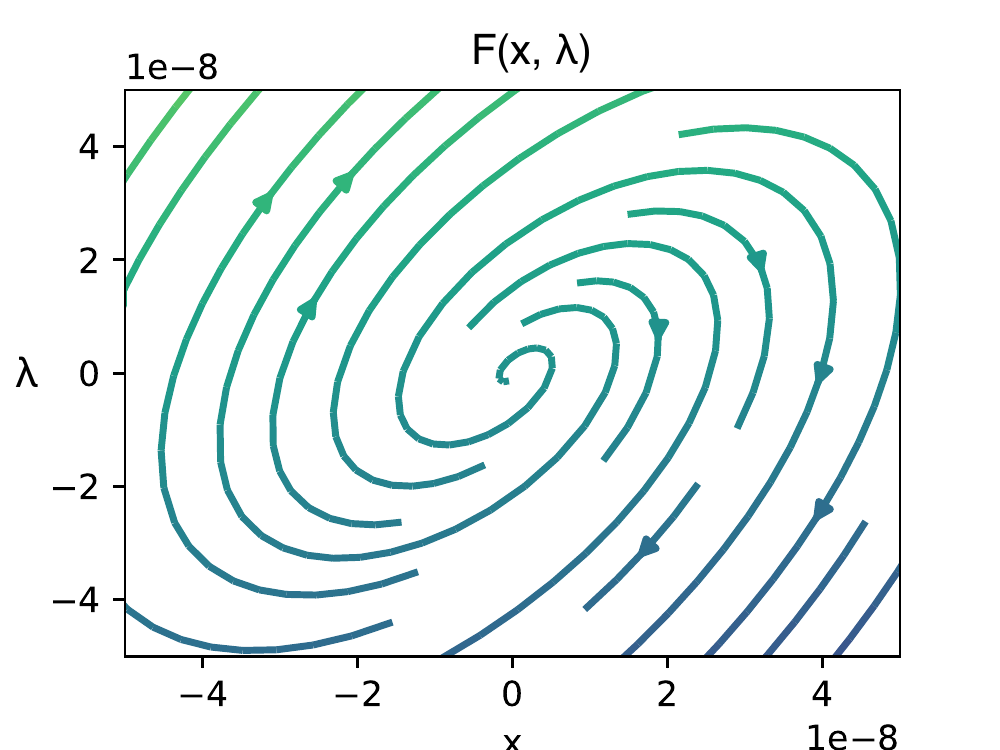}
    \includegraphics[scale=0.4]{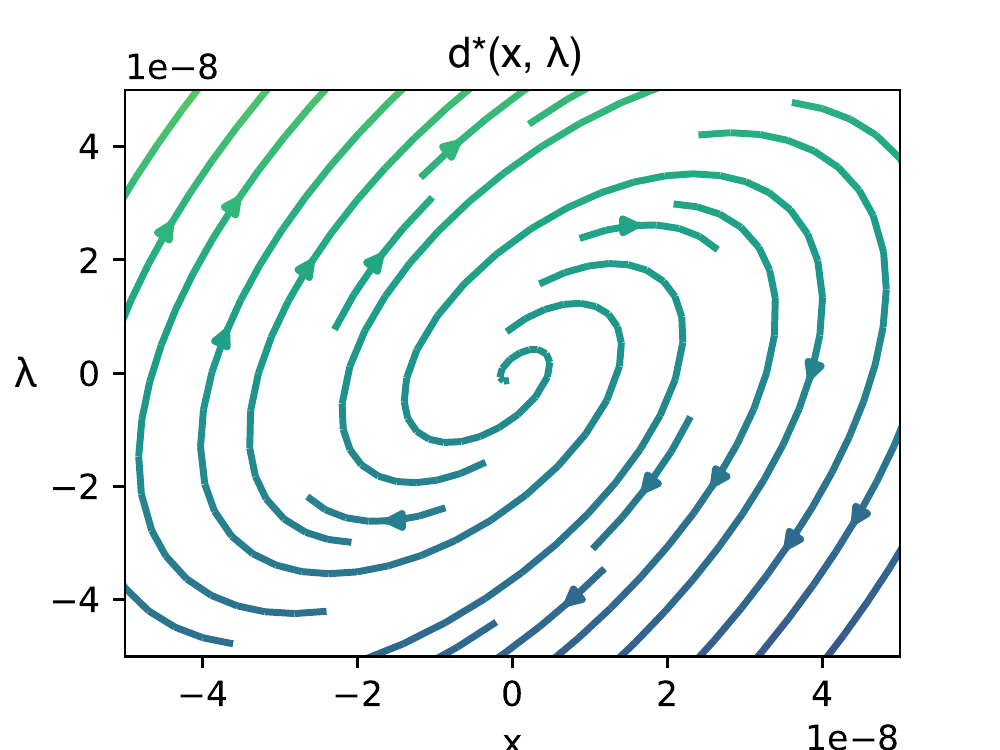}
    \caption{\textit{Left:}  The thermodynamic force field (\textit{streamline plots}) from theory for the same parameter choice as in Figure 1b of the main text. \textit{Right:} The optimal $d^* \propto \mathcal{F}$ obtained from the algorithm for the same parameter choice.}
    \label{fig:12p5}
\end{figure}

Since we have used a finite amount of data to construct the cost function, it will be prone to statistical errors. Therefore we independently maximise the cost function for different 12.5s long data sets, and take their mean value as the optimized estimate of $\sigma$. We show the value of sigma inferred ($\sigma_L$) as a function of the number of steps in the optimization algorithm for different 12.5s long data sets in Fig.\ \ref{fig:12p5F}.
With the numerical data, we also find that the optimal field $d^*$ (see Eq.\ \eqref{eq:psi}) is proportional to the thermodynamic force field $\mathcal{F}$ (Eq.\ \eqref{eq:optf}). We demonstrate this in Fig.\ \ref{fig:12p5}.

\subsubsection{Comparison of the TUR and the SFI inference schemes}

In this section, we compare the TUR and SFI based inference algorithms with respect to the sampling rate $\Delta t $, (for a fixed total length of the trajectory $\tau = 1s$), as well as the length of the single trajectory used for inference $\tau$ (with a fixed $\Delta t=0.0001s$). All the results are obtained for numerically generated data.

\begin{figure}[H]
    \centering
    \includegraphics[scale=0.45]{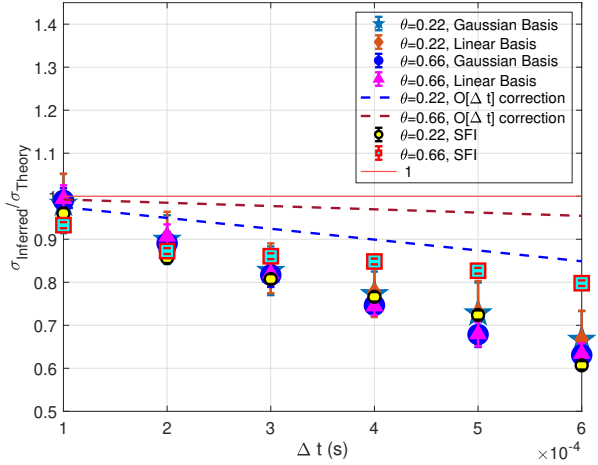}
    \caption{ Inference as a function of $\Delta t$. The dashed lines are the theoretical values corresponding to the lowest-order correction in $\Delta t$ from Eq. (\ref{correctiondt}). The points correspond to the ratio of inferred $\sigma$ value to the theoretical $\sigma$ value, using Linear and Gaussian basis in the TUR inference scheme as well as with a Linear basis in the SFI scheme, for two $\theta$ values. The error bars in the TUR based inference scheme correspond to averages over eight independent realizations of length $12.5s$ sampled at a time interval $\Delta t$. The error bars in the SFI scheme correspond to a self-consistent estimate of the inference error that the algorithm provides for the whole 100s trajectory, for every $\Delta t$. }
    \label{fig:shortlimit}
\end{figure}

In Fig.\ \ref{fig:shortlimit}, we demonstrate, how much the inferred value can differ from the true entropy production rate, if the sampling time-step of the trajectory is increased, for the TUR inference scheme in the Linear and Gaussian basis and the SFI scheme in the Linear basis. Results show that, if we use larger $\Delta t$ values, the inferred value can become significantly lower, as much as $2/3rd$ of the original value. We also find that when $\theta = 0.66$, SFI performs slightly better as compared to the TUR scheme. 

\begin{figure}[H]
    \centering
    \includegraphics[scale=0.45]{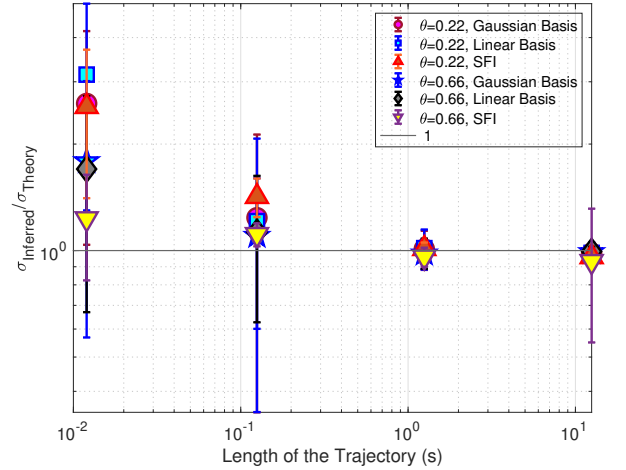}
    \caption{ {Inference as a function of the Length of the trajectory in seconds, with $\Delta t=0.0001s$. Again, the points correspond to the two basis functions, two different $\theta$ values as well as the results of the SFI algorithm for the two $\theta$ values. The error bars are computed in the same way as in Fig.\ \ref{fig:shortlimit}.} }
    \label{fig:sizeofdata}
\end{figure}

\begin{figure}[H]
    \centering
    \includegraphics[scale=0.45]{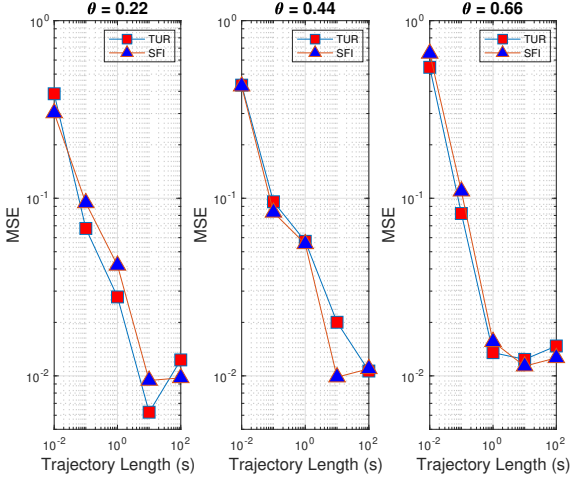}
    \caption{The Mean-Squared-Error (MSE) in the linear projection coefficients $C$ of the thermodynamic Force field $\mathcal{F}^\mu(\textbf{x}) = C^{\mu\nu}x^\nu $ defined as $MSE=\frac{\sum \left(\hat{C}^{\mu\nu}-C^{\mu\nu}\right)^2}{\sum (C^{\mu\nu})^2}$, as a function of the length of the steady state trajectory used for inference. The red squares correspond to the short-time inference scheme and the blue triangles correspond to the SFI scheme. Both schemes perform similarly in all cases.}
    \label{fig:Ferr}
\end{figure}

In Fig.\ \ref{fig:sizeofdata}, we plot the estimation results of our scheme as well as the SFI scheme for $\theta = 0.22$ and $\theta = 0.66$. 
The analysis shows that at least $10^4$ data points ($\sim 1s$ data with $\Delta t =0.0001$) are required to get a reliable estimate of the entropy production rate, using the short-time inference scheme. We note that this is well within the capacity of current experiments \cite{kumar2020exponentially}. We also notice that, when the amount of data is less ($10^3$ data points $\sim 0.1s$ with $\Delta t =0.0001s$), the SFI based inference scheme performs better.
Finally, we demonstrate that our TUR-based short-time inference scheme for the thermodynamic force field, from a single trajectory of a given length, is equivalent to the SFI based approach for computing the thermodynamic force field. We quantify this by using the Mean-Squared-Error (MSE) function defined as $MSE=\frac{\sum \left(\hat{C}^{\mu\nu}-C^{\mu\nu}\right)^2}{\sum (C^{\mu\nu})^2}$, where $\hat{C}$ and $C$ are the inferred and the theoretical projection coefficients of the thermodynamic force field in the linear basis, such that $\mathcal{F}^\mu(\textbf{x}) = C^{\mu\nu}x^\nu $. We find that both the schemes perform equally good, as the errors decrease similarly with the trajectory length, as shown in Fig.\ \ref{fig:Ferr}. The results are shown for three different values of $\theta$. 

 In conclusion, we find that when the trajectory length is $\sim 1s$ with sampling rate $\Delta t =0.0001$, both the SFI and TUR-based schemes are equally good and give robust estimates of the entropy production rate, as well as the thermodynamic force field. If the trajectory length is $\sim 0.1s$ with sampling rate $\Delta t =0.0001$, SFI is seen to perform marginally better in estimating the entropy production rate. Similarly, for a given length of the trajectory, with a fixed, but lower sampling rate $\Delta t$, we find that SFI performs better for a larger $\theta$ value. It would be very interesting to investigate these results further for systems with non-linear forces and space-dependent diffusion terms.

\subsubsection{Hyper-parameter tuning for the Particle-Swarm optimizer}
\begin{figure}[H]
    \centering
    \includegraphics[scale=0.45]{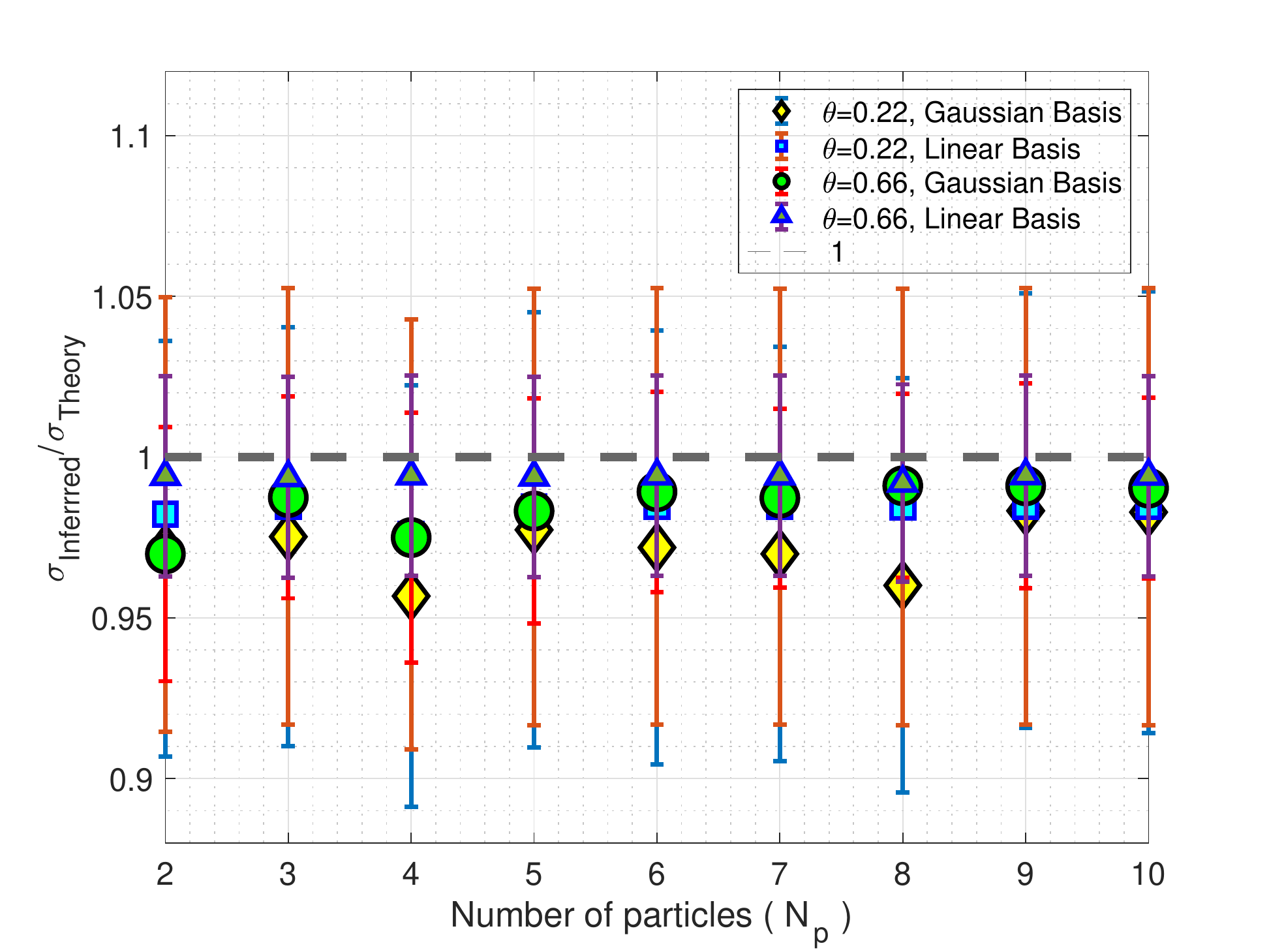}
    \caption{ Inference as a function of the number of particles used for the optimization for the Gaussian and Linear basis as well as different $\theta$ values.} 
    \label{fig:Np}
\end{figure}

\begin{figure}[H]
    \centering
    \includegraphics[scale=0.45]{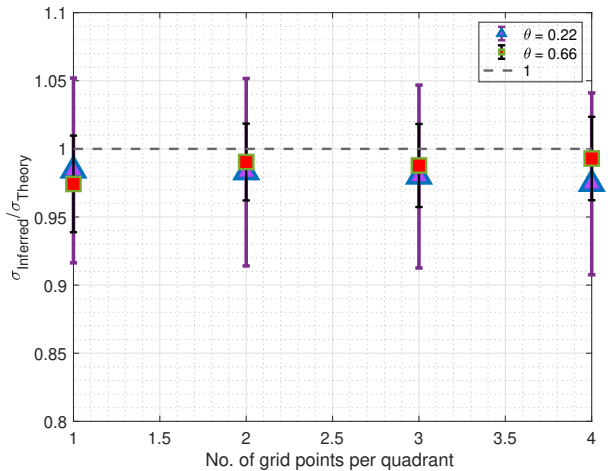}
    \caption{Inference as a function of the number of grid points  used for inference in the Gaussian basis, in the $x$ (same in the $\lambda$) direction.} 
    \label{fig:Ng}
\end{figure}

The specific algorithm we have used in this study is built upon the Particle swarm optimizer. We note that this choice is not very crucial, and Refs. \cite{Shun:eem} and \cite{van:epe} contain equivalent algorithms that can be used to perform the optimization task. In the algorithm we present here, the optimization is done using both a Linear and Gaussian basis, and as seen in Fig.\ \ref{fig:shortlimit} and Fig.\ \ref{fig:sizeofdata}, the choice of the basis function is not so crucial  for the evaluation of $\sigma$ for this problem with linear forces.  We further looked at the effect of one of the crucial hyper-parameters in the algorithm, which is the Number of particles ($N_p$). As shown in Fig.\ \ref{fig:Np}, we find that the effect is negligibly small for the Linear basis, and the inferred value of $\sigma$ remains more or less the same, independent of $N_p$. For the Gaussian basis, we find that  the estimation improves with the number of particles. We have used $N_p = 10$, for most of our simulations using the Gaussian and the Linear basis. 

Yet another hyper-parameter relevant for inference, when we use the Gaussian basis, is the number of grid points ($N_g$) that we choose in one quadrant along both $x$ and $\lambda$ directions. In this case, the number of grid points determine the number of parameters in the optimization problem ( = numbers of Gaussian functions in the basis), and while more of them can slow down the algorithm, too few can lead to the thermodynamic force not being well resolved. In this work, we choose $N_g^x = N_g^\lambda$ in one quadrant, and therefore the total number of grid points are  $M= 4 \times N_g^x \times N_g^\lambda$. The plot below shows the dependence of the inference on the number of grid points. We plot the ratio of inferred $\sigma$ vs. the number of grid points in the $x$ (same in the $\lambda$) direction, for $\theta = 0.22$, $\theta = 0.66$. We note that the inference works slightly better when $N_g^x = N_g^\lambda = 2$. We have therefore used $N_g^x = N_g^\lambda = 2$ in this work. 


\newpage
\begin{widetext}
\section{Supplemental Information}
This material contains supplemental information for the results presented in the manuscript "Quantitative analysis of non-equilibrium systems from short-time experimental data".

\section{I. The Stochastic Force Inference technique (SFI)}
In this work, we have used the Stochastic force inference technique, introduced in \cite{frishman:sfi} for bench-marking our results for both experimental setups. In the SFI technique, both the drift and diffusive terms of the stochastic equation are represented using an appropriate finite set of basis functions, and then the coefficients are inferred from the trajectory using projective techniques. Using these estimates of the drift and diffusion terms, one can also obtain thermodynamic quantities such as the entropy production in the stationary state. In this work, we used the SFI technique (in the linear basis) to associate a dynamical equation to the experimental trajectory data used in Fig. 2, and to obtain an independent estimate of the entropy production rate which then can be compared with the results from the short-time inference scheme. In these cases, the conversion factor for the trajectory data is fixed such that the $x$ component of the diffusion matrix has the correct room temperature value. For the case $\theta = 0.22$, we obtain, for one $100 s$ trajectory and $\Delta t = 0.0001 s$, the drift and diffusion matrix to be,
\begin{align}
\label{th22}
    \textbf{F}&= \left( \begin{array}{cc}
         -303.21  &622.19\\
  1.79& -419.45
    \end{array}\right),& \textbf{D}&= \left(\begin{array}{cc}
       1.645\times 10^{-13}&0\\
   0& 3.75 \times 10^{-14}
    \end{array}\right)
\end{align}

For the case, $\theta = 0.66$, we obtain the  drift and diffusion matrix to be,
\begin{align}
\label{th66}
    \textbf{F}&= \left( \begin{array}{cc}
         -515.79&  696.86\\
  4.60& -417.56
    \end{array}\right),& \textbf{D}&= \left(\begin{array}{cc}
        1.65\times 10^{-13} & 0\\
 0 & 1.11\times 10^{-13}
    \end{array}\right)
    \end{align}
The diffusion matrix agrees well with the theoretical model we wanted to realize in experiment,  by construction. The $x$ component is fixed such that it agrees with the room temperature value (which gives us the conversion factor for the $x$ - data ) of the Diffusion constant. The $\lambda$ component of the Diffusion is set externally by the Ornstein-Uhlenbeck process used to control  the mean position of the optical trap. The same applies to the $\lambda$ component of the force matrix \textbf{F}. On the other hand, the $x$ component of the Force matrix, is determined by the effective force the trapped particle experiences along the $x$ direction.  For both the $\theta$ values, the theoretical \textbf{F} matrix is supposed to be,
\begin{align}
    \textbf{F}_{Theory}= \left( \begin{array}{cc}
         -848.23&  848.23\\
  0& -400
    \end{array}\right),
    \end{align}
for $f_c = 135\; Hz$ and $\tau_0 = 0.0025s$. As we see from Eq.\ \eqref{th22} and Eq.\ \eqref{th66}, the Forces constructed from the trajectory data is different from the theoretical model. We think that the differences may be mainly due to the long  experimental stretches we used in this study, during which the particle might have diffused along $y$ or $z$ directions in the trap, where it experiences a weaker confinement along the $x$ direction. In this work however, we took into account this error, by directly comparing the results from inference with the results obtained from the SFI technique. We further computed an effective theoretical expression for the entropy production rate by taking the average \textbf{F} matrix over the six points in Fig. 2. We first obtain, 
\begin{align}
\label{fm}
    \textbf{F}=\left(\begin{array}{cc}
        -440.31\pm 91.17 &663.24\pm 49.9  \\
       2.67\pm 1.48  & -415.17\pm 3.73
    \end{array}\right)
\end{align}
Since it is a Linear model, we can still solve it and obtain an analytical expression for $\sigma$, as well as the thermodynamic Force Field using standard techniques \cite{Gingrich:qua}. We get,
\begin{align}
    \sigma=-\frac{(D_{22} F_{12}-D_{11}F_{21})^2}{D_{11} D_{22} (F_{11}+F_{22})}
\end{align}
The thermodynamic force field is given by, 
\begin{align}
    \mathcal{F}=\left(\begin{array}{c} \frac{(D_{22} F_{12}-D_{11} F_{21}) (D_{11} F_{22} \lambda (F_{11}+F_{22})+D_{11} F_{21} (F_{22} x-F_{12} \lambda )+D_{22} F_{12} (F_{11} x+F_{12} \lambda))}{D_{11} \left(D_{11}^2 F_{21}^2+D_{11} D_{22} \left(F_{11}^2+2 F_{11} F_{22}-2 F_{12} F_{21}+F_{22}^2\right)+D_{22}^2 F_{12}^2\right)}\\-\frac{(D_{22} F_{12}-D_{11} F_{21}) \left(D_{11} F_{21} (F_{21} x+F_{22} \lambda)+D_{22} \left(F_{11}^2 x+F_{11} F_{12} \lambda+F_{11} F_{22} x-F_{12} F_{21} x\right)\right)}{D_{22} \left(D_{11}^2 F_{21}^2+D_{11} D_{22} \left(F_{11}^2+2 F_{11} F_{22}-2 F_{12} F_{21}+F_{22}^2\right)+D_{22}^2 F_{12}^2\right)}
    \end{array}\right)
\end{align}
Taking the \textbf{F} matrix elements from Eq.\ \eqref{fm} and using $D_{11} = 1.645 \times 10^{-13}$ and  using $\theta =\frac{D_{22}}{D_{11}}$, we obtain, 
\begin{align}
    \sigma(\theta)= -4.13934\, +513.929 ~\theta +\frac{0.00833487}{\theta }.
\end{align}
We use these expressions as the SFI theory model in Fig. 2 to obtain the theoretical estimate of $\sigma$ for the experimental data, and as the theoretical Force field form used for comparison in Fig. 3. The red dashed lines in Fig. 2 account for the error bars in the terms in the \textbf{F} matrix in Eq.\ \eqref{fm}. Notice that, the expression for $\sigma$ diverges in the $\theta \rightarrow 0$ limit, as opposed to the case in Eq.\ \eqref{eq:sigmaF}. This is due to having a small non-zero $F_{21}$ term in the Force matrix. If we treat this term to be negligible and set it to 0, the divergence goes away.

Here we would also like to point out one advantage of describing a non-equilibrium system in terms of the entropy production rate $\sigma$.  The advantage is that $\sigma$ does not change even if we multiply the individual components of a trajectory using a scaling factor. 
On the other hand, information such as the force acting on the system or the Diffusion constant, will be modified and scaled if the coordinates are transformed. In our case, we use this to our advantage when studying the case with the bubble (Figs. $5$ and $6$), where the exact conversion factors for the trajectory data are hard to obtain. 

\section{II. The heat dissipated in the medium for the case with the bubble}
 In this work, we have obtained an estimate for the average total entropy production of
a colloidal particle maintained in a steady state by being confined in a shaken trap (the stochastic sliding parabola model), under
two different experimental conditions, namely without and with a microscopic bubble in the vicinity of the trap. 
The average total entropy production for a system in steady state is also
the same as the heat dissipated by the system into the surrounding bath
(at constant temperature $T$). This heat dissipated includes the heat associated with keeping the system in a steady state (by shaking the trap)
and, if there is a flow, the heat associated with the work done by the flow on the particle. As we argue below, the latter component cannot be
obtained by the short-time inference scheme, and is related to a fundamental limitation of the applicability of the TUR \cite{Seifert:inf} related to how the flow term is dealt with. 


We begin with a possible generic form of the Langevin equation in the presence of the bubble, 
\begin{align}
\label{eq:sspmodified}
    \dot{x}-u_d &= -\frac{\left(x-\lambda \right)}{\tau_d}+\sqrt{2D_d}\;\eta(t),\\
    \dot{\lambda}&=-\frac{\lambda}{\tau_0}+\sqrt{2A}\;\xi (t),
\end{align}
where $u,\;\tau$ and $D$ are taken to be slowly varying functions of $x$, and essentially treated as constants ($u_d,\;\tau_d$ and $D_d$)  at a distance $d$ from the bubble, where the particle is trapped.

First, we notice that under the transformations $x \rightarrow x^\prime = x- \tau_d u_d$, the above equations  map to the Stochastic sliding parabola model, with the parameters $\tau=\tau_d$ and $D=D_d$. This observation also demonstrates that for the above system, the mean position of the particle is no longer at the center of the trap, but is instead $\left \langle x \right\rangle = u_d \tau_d $.  Now we look at the entropy production in this system, using the standard definitions in Stochastic thermodynamics.

Since the system is in a stationary state, the actual rate of entropy production can be obtained in terms of the heat ($q$) dissipated to the medium at a temperature $T$ as,
\begin{align}
    \sigma=\frac{q}{T}.
\end{align}
 However there is an ambiguity on how to obtain the correct value of $\sigma$, arising from two choices of transformations for the flow term under time-reversal \cite{Seifert:ref}.

The first approach is to let the flow term reverse it's sign under time-reversal, as physically meaningful for a velocity variable. 
This gives an estimate of medium entropy production \cite{Seifert:ref} as,
\begin{align}
\begin{split}
\label{eq:former}
    \sigma&=\frac{q}{T}\\ &= \frac{\langle (\dot{x}-u_d)(-\nabla_x V)\rangle}{T}\\ &= \frac{\langle (\dot{x}-u_d)(\lambda-x)\rangle}{T}.
     \end{split}
\end{align}
The observed trajectories of the colloidal particle, on the other hand, only show the effect of the flow $u_d$ as a constant external force acting on the system, which only amounts to shifting the mean position of the colloidal particle in the direction of the flow. This leads to a second (naive) approach to the entropy production in this system as,
\begin{align}
\begin{split}
\label{eq:latter}
    \sigma^\prime  &= \frac{\langle \dot{x}(-\nabla_x V+u_d\tau_d)\rangle}{T}\\&=\frac{\langle \dot{x}(\lambda-x+u_d\tau_d)\rangle}{T}.
    \end{split}
\end{align}

The physical distinction between the two definitions is as follows: 
when there is a background flow in the medium, this flow has to constantly do work against the confining potential to maintain the particle in it's "new" average position.This is an additional contribution to entropy production, that is only accounted for in the definition in Eq.\ \eqref{eq:former}. In other words, the particle trajectories do not carry information about this and hence the short-time inference scheme, which is based on TUR and the information carried by particle trajectories, only predicts the quantity $\sigma^\prime$ in Eq.\ \eqref{eq:latter}.
$\sigma$ and $\sigma^\prime$ are related by,
\begin{align}
\begin{split}
    \sigma &= \sigma^\prime +\frac{u_d^2 \tau_d}{T},\\
    &\geq \sigma^\prime.
    \end{split}
\end{align}
When the flow velocity $u_d=0$, they are the same. 
\section{Currents in the non-equilibrium stationary state}
Systems in a non-equilibrium stationary state are characterized by a non-vanishing current in the phase space \cite{battle2016broken}. For the colloidal system we consider, these currents can be estimated from the trajectory data as,
\begin{align}
\label{eq:defc}
    \left[ \begin{array}{c}
         J_{x}(x,\lambda)  \\
         J_\lambda (x,\lambda)
    \end{array}\right] &=  \Bigg[ \Bigg \langle \begin{array}{c}
         x(t+\Delta t) -x(t)  \\
         \lambda(t+\Delta t)-\lambda(t)
    \end{array}\Bigg \rangle_{x,\lambda}\\&-\Bigg \langle \begin{array}{c}
         x(t) -x(t-\Delta t)  \\
         \lambda(t)-\lambda(t-\Delta t)
    \end{array}\Bigg \rangle_{x,\lambda}\Bigg]\frac{P_{ss}(x,\lambda)}{2\Delta t}.
\end{align}
Using Eq.\ \eqref{eq:defc} we estimate currents in the case when the bubble is present in the vicinity of the optical trap. We find that the phase space currents are reduced in magnitude. We demonstrate this with surface plots of the two components of the currents in Fig.\ \ref{fig:bubcc} for the case discussed in Figure 5 in the main text.
\begin{figure}[H]
    \centering
    \includegraphics [scale=0.4]{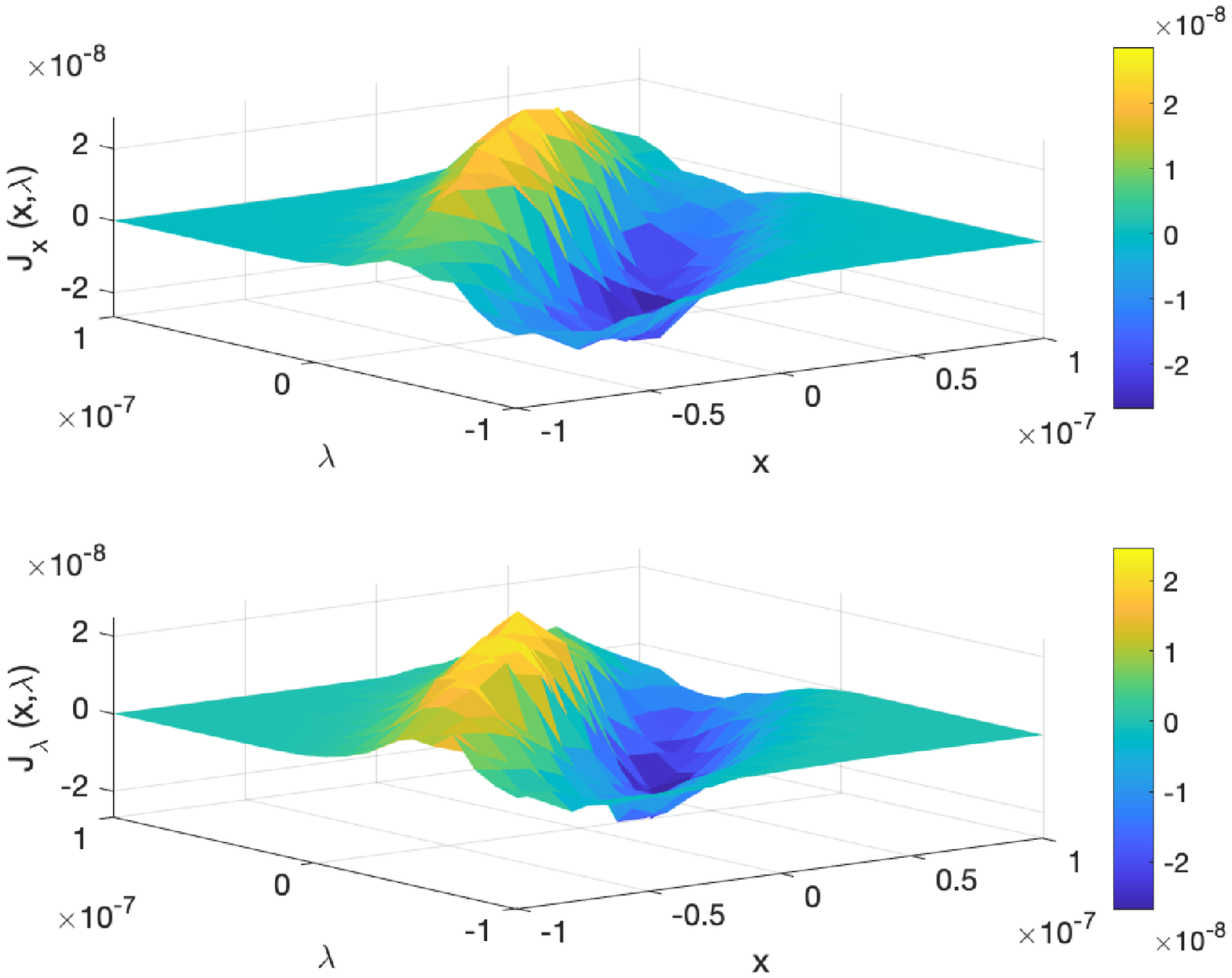}
     \includegraphics [scale=0.4]{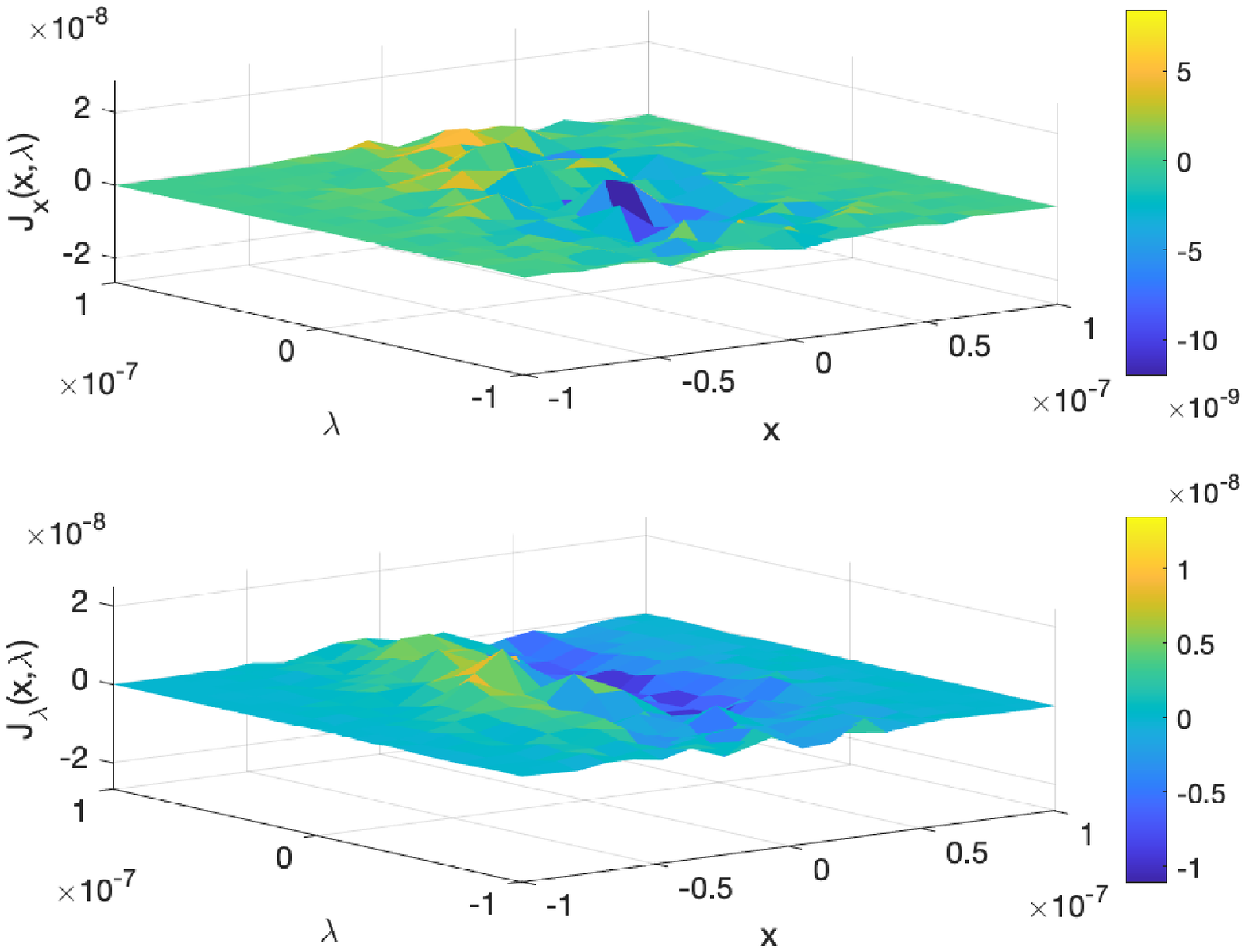}
    \caption{Surface plots of the two components of the currents ($J_x$ and $J_\lambda$) (Eq.\ \eqref{eq:defc}) for the case discussed in Figure 5 of the main text. \textit{Left:} Case without the bubble in the vicinity of the optical trap. \textit{Right:} Case with the bubble in the vicinity of the optical trap. We find that the magnitude of the currents are reduced in the vicinity of the bubble.}
    \label{fig:bubcc}
\end{figure}
\section{Parameter values}
\paragraph{Figure 1: } $\tau=\frac{1}{2 \pi\; f_c}=0.0012$, $\tau_0 =0.0025$, $D =1.6452\times 10^{-13}$, $A= [0.1,\;0.15,\;0.2,\;0.25,\;0.3,\;0.35]\times (0.6\times 10^{-6})^2 $.
\paragraph{Figure 2: } $f_c=135\pm 10$, $\tau_0 =0.0025$, $D =1.6452\times 10^{-13}$, $A= [0.1,\;0.15,\;0.2,\;0.25,\;0.3,\;0.35]\times (0.6\times 10^{-6})^2 $. 
\paragraph{Figure 5: } $f_c=57 \pm 3\; Hz$, $\tau_0 =0.025$, $D =1.6452\times 10^{-13}$, $A= 0.3\times (0.6\times 10^{-6})^2 $. 
\paragraph{Figure 6: } $f_c=135 \pm 10\; Hz$, $\tau_0 =0.0025$, $D =1.6452\times 10^{-13}$, $A= 0.3\times (0.6\times 10^{-6})^2 $. 
\end{widetext}

\end{document}